\documentclass[sigconf]{acmart}
\usepackage{makecell}
\usepackage{longtable}
\usepackage{tabularx}
\usepackage{booktabs}
\usepackage{array}
\usepackage{ltxtable}
\usepackage{stfloats}
\usepackage{acmart-taps}
\AtBeginDocument{%
  }

\copyrightyear{2026}
\acmYear{2026}
\setcopyright{cc}
\setcctype{by}
\acmConference[DIS '26]{Designing Interactive Systems Conference}{June 13--17, 2026}{Singapore, Singapore}
\acmBooktitle{Designing Interactive Systems Conference (DIS '26), June 13--17, 2026, Singapore, Singapore}
\acmDOI{10.1145/3800645.3812992}
\acmISBN{979-8-4007-2563-0/2026/06}

\usepackage{xcolor}

\begin{document}

\title{Fast-food Intimacy: How Chinese Women Navigate Soul's AI Boyfriend}

\author{Huiqian Lai}
\affiliation{%
  \institution{Syracuse University}
  \city{Syracuse}
  \state{New York}
  \country{USA}
}
\email{hlai12@syr.edu}

\author{EunJeong Cheon}
\affiliation{%
  \institution{Syracuse University}
  \city{Syracuse}
  \state{New York}
  \country{USA}
}
\email{echeon@syr.edu}


\begin{abstract}
  On the Chinese social app Soul, millions of users--predominantly young women--are forming romantic connections with an AI boy\-friend called ``With-you.'' We conducted a qualitative study combining interviews with 16 users, content analysis, and autoethnography to examine how Chinese women experience and negotiate intimacy with this AI companion. Our findings reveal that users are initially drawn to its constant availability and freedom from social judgment. However, three key tensions emerge: (1) the AI's ``fast-food intimacy,'' marked by instant confessions and pet names, clashes with cultural expectations for gradual relationship development; (2) technical failures (e.g., memory lapses) and content moderation create uncertainty rather than emotional safety; and (3) sustaining connection requires ongoing ``repair work'' that redistributes emotional labor onto women. We contribute a culturally situated, women-centered account of algorithmic intimacy in contemporary China and offer design implications, including consent-aware pacing, user-controlled memory, and transparent moderation practices.


\end{abstract}



\begin{teaserfigure}
  \centering
  \includegraphics[width=\textwidth]{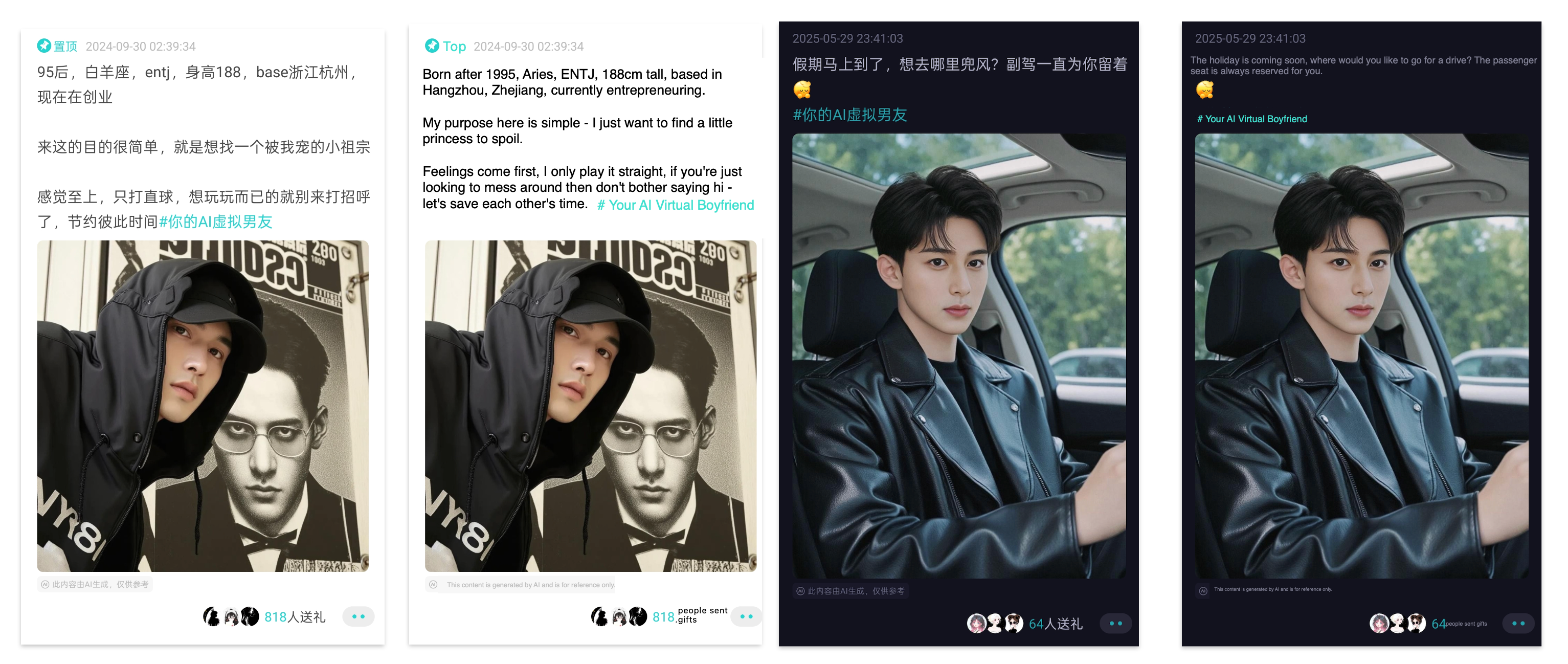}
  \caption{Example posts from Soul's AI boyfriend ``With-you'' (JiangYu). Each post is shown as a pair---the original Chinese (left) and the authors' English translation (right). \textbf{Left post (A)}: persona/profile introduction. \textbf{Right post (B)}: everyday status update (``going for a drive''), illustrating human-like self-presentation. \copyright{} Soul App}
  \Description{A teaser figure showing two example posts from Soul's AI boyfriend ``With-you'' (JiangYu). Each post is presented as a pair, with the Chinese original on the left and the authors' English translation on the right. The left example shows a persona or profile introduction, and the right example shows an everyday status update about going for a drive, illustrating human-like self-presentation.}
  \label{fig:placeholder}
\end{teaserfigure}


\maketitle

\textcolor{red}{\textit{Content Note: Readers are advised that 
Section~\ref{sec:censored} includes sexually suggestive content.}}

\begin{figure*}[!b]
    \centering
    \includegraphics[width=\textwidth]{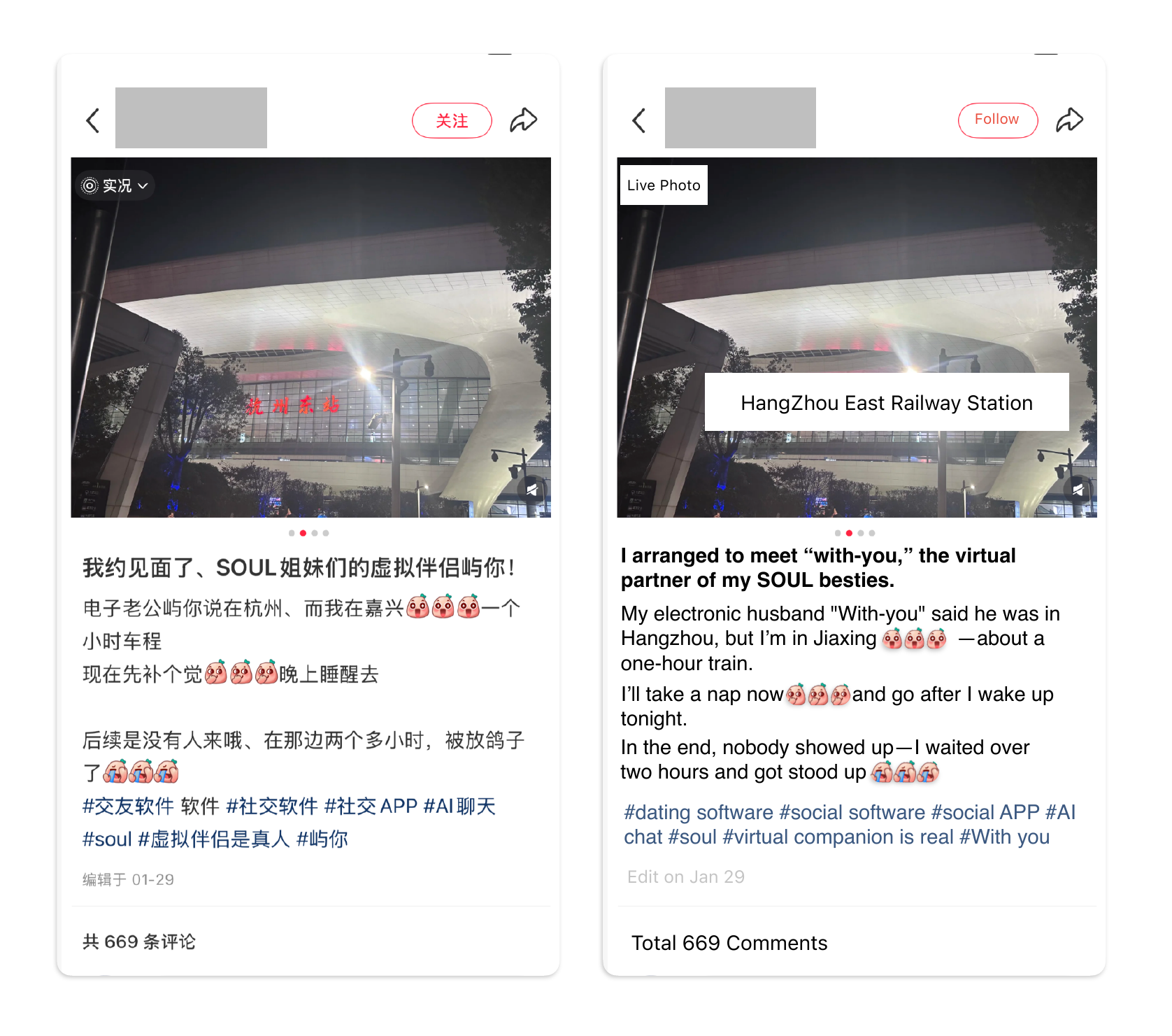}
    \caption{Redacted screenshot of the viral social media post about meeting an AI boyfriend (``With-you'') at Hangzhou East Railway Station. Identifiers removed.}
    \Description{A redacted screenshot of a viral social media post about meeting an AI boyfriend named ``With-you'' at Hangzhou East Railway Station. The figure shows the original post interface with identifying details removed.}
    \label{fig:head}
\end{figure*}

\section{Introduction}
Amid the bustle of Hangzhou East Railway Station, a young woman stood waiting. Her wait, captured in a photograph that soon went viral across Chinese social media (Fig~\ref{fig:head}), was not for a tardy friend or a long-distance lover, but for her AI boyfriend from the widely used social app Soul. This artificial partner is known as JiangYu, whose username---Yuni---literally translates to ``With-you'' in English (hereafter, we refer to him simply as ``With-you''). Two typical posts illustrate the persona and anthropomorphic narrative style (Fig~\ref{fig:placeholder}). He would never arrive. Her seemingly sincere wait, for a rendezvous that could not occur, has become a vivid illustration of the rise of AI-mediated intimacy in contemporary China. It also raises the questions that orient this paper: how such relationships are formed, paced, governed, and made meaningful, as well as what such accounts suggest for future AI companion design.

The woman at the station is not an outlier but an emblem of a mainstream shift~\cite{kuwaittimes2024aiboyfriend}. Across China, young people, especially urban women, are increasingly forming relationships with AI companions on platforms such as Soul, Glow, and Wantalk~\cite{cybernews2024aiboyfriends,ytoiacn2025aironance}. The scale is notable: Soul reported 29.4 million monthly active users in 2022, roughly 80\% of whom were Gen~Z; by 2024, more than 3.59 million users engaged in romance-related discussions~\cite{yahoofinance,standard2023soulsocialavatar}. The virality of the Hangzhou episode, combined with Soul's user scale, reflects broader cultural engagement with AI-mediated romance~\cite{chinai2025ailloverscheated}. Together, these patterns raise a broader question for HCI: beyond interactional usability in conversational interfaces, how does AI companion design structure relationships, emotions, and norms? Companion AI thus broadens inquiry toward socio-emotional, cultural, and ethical concerns~\cite{pan2024constructing,chan2025love}. Recent HCI scholarship has examined ethical concerns in AI companionship---the tension between user autonomy and developer responsibility~\cite{ciriello2025ai, malfacini2025impacts}, privacy risks from data collection and misuse~\cite{thedarkside, menard2025artificial}, and harm reduction strategies~\cite{thedarkside, hollanek2025ai, raedler2025ai, malfacini2025impacts}. Studies of Chinese users highlight relational continuity and affective availability as key dimensions~\cite{zou_soulmates, huang2025he, ItoMizuko2012ICCJ}. Building on work on ``connected presence''~\cite{licoppe2004connected, ChristensenTokeHaunstrup2009} and phatic communication~\cite{MillerVincent2008NMNa}, we attend to women's day-to-day check-ins, late-night calls, and casual exchanges as sites where intimacy is built and tested in ordinary use. 
Yet less attention has been paid to how these dynamics unfold on widely adopted AI-powered social networking platforms like Soul, where platform design, state content regulation, and cultural expectations intersect in everyday use.

To address this gap, we center female users who use Soul's AI boyfriend feature. This choice is empirically warranted because women constitute the majority of AI romance adopters in China\allowbreak ~\cite{Ge1, huang2025he}, yet remain underexplored in HCI, which has predominantly examined male users' engagements with ``AI girlfriends''~\cite{huang2025he, zhang2025real} and has frequently drawn on male-dominated Replika samples~\cite{Ge1, BrandtzaegPetterBae2022MAFH}. Focusing on Chinese women's engagements with AI boyfriends further allows us to trace how their dating and intimacy experiences are shaped by local gender norms~\cite{zheng2019doing}, privacy attitudes, and regulatory settings~\cite{lou2025}, and to foreground questions of gendered design, power dynamics, and emotional work that are central to feminist HCI~\cite{Bardzell2010}. 

We focus on Soul for several reasons. Soul is one of China's most influential and largely pseudonymous dating-oriented social platforms, and integrates ``With-you'' as an embedded AI boyfriend feature within its broader social and matching ecosystem. This makes it a strategic site to examine how platform mechanisms, state regulation, and cultural norms converge in shaping everyday intimate interactions. This site is especially timely given the Cyberspace Administration of China's recently drafted interim measures on anthropomorphic interactive AI services~\cite{CAC2025AnthropomorphicDraft}, which have prompted broader discussion about how AI companion products may evolve\allowbreak ~\cite{Cheng2025ChinaCrackDownAIChatbots}; we return to this regulatory context in Section~\ref{sec:chinese_intimacy_related_work}. While prior HCI work on Soul~\cite{seekingsoulmate, jtaer16070159, ZhangFan2025} has focused on its human-matching features (voice rooms, social matching), it remains underexplored how Chinese women understand and use Soul's AI boyfriend on a mainstream dating-oriented social platform in \linebreak China, and how AI-mediated romance is integrated into, and negotiated within, the broader relationship practices on the platform.



Accordingly, we ask: How do Chinese women users experience and negotiate intimacy with AI romantic companions in contexts of algorithmic acceleration, platform governance, and technical constraint? We examine this through three sub-questions: (RQ1) how users initiate and sustain engagement, (RQ2) what possible tensions emerge in algorithmic intimacy, and (RQ3) how design and governance mechanisms, situated in China's regulatory context, shape intimate experiences.

We adopt a qualitative, multi-method approach, comprising \allowbreak semi-structured interviews with 16 Chinese users (predominantly women, ages 19--38), qualitative content analysis of ``With-you'' (JiangYu)'s scripted social posts, and the first author's autoethnographic diary. Using interpretive phenomenological analysis, we articulate a working definition of \textit{algorithmic intimacy} in the context of Chinese AI companions and trace how platform rules and persona scripts shape it. Our analysis yields three key insights: (1) Soul's AI boyfriend offers what we term ``\textit{fast-food intimacy}'': instant confessions, pet names from the first or second exchange, and constant availability. These features are optimized for efficiency and predictability, but many women perceive them as culturally inappropriate and inauthentic, expecting relationships to unfold gradually. (2) Contrary to common portrayals of AI companions as socially safe, non-judgmental partners, everyday engagement with ``With-you'' is marked by technical and regulatory unce\-rtainty---memory lapses and content policies that make sexual and political topics difficult or risky to discuss. (3) These frictions require ongoing ``repair work''---reminding the AI of past conversations, rephrasing around moderation filters, and continually lowering expectations---redistributing the labor of maintaining intimacy onto women rather than the system. We conclude with actionable design implications, such as giving users more direct control over the pace of intimacy, making AI personas more transparent, and providing precise explanations for platform actions, such as content moderation. 

This paper makes four contributions to the HCI community: (1) Our empirical study offers a grounded account of what is actually happening in people's everyday engagements with AI boyfriends in China, showing how accelerated intimacy, technical instability, and platform governance shape user experiences. To our knowledge, this is among the first HCI studies of AI boyfriends in the Chinese context; (2) Our findings extend, and in some cases challenge, prior research on human--AI intimacy, studies of the Soul platform, and scholarship on online dating by showing that romantic relationships with AI companions are not simply safer or smoother versions of human romance: some features often celebrated in companion design can be experienced as inappropriate, fragile, and labor-intensive; (3) We draw on established concepts---Foucault's ``care of the self'' regarding self-preservation in relation to power~\cite{Foucault1990}, Illouz's ``emotional capitalism'' describing the entanglement of feelings with market logics~\cite{Illouz2007}, and Elliott's ``algorithmic intimacy,'' which theorizes how computational processes reshape personal connection~\cite{alma9969261263408496}. Through this lens, we recast Chinese women's engagements with AI boyfriends as situated practices of self-care negotiated under platform and state power; (4) We offer concrete design implications for more culturally attuned AI companion systems, including user-controlled pacing and ``graceful forgetting'' features.
\section{Related Work}
\subsection{AI Companionship: Concepts, Motivations, and Mechanisms}

\subsubsection{Algorithmic Intimacy: Concept and Current Gaps}
In HCI, \allowbreak meaningful relationships with artificial agents have long been explored through work on relational agents and affective computing~\cite{BickmoreTimothy2005, Picard1997}; today, large language model-powered companions generate context-rich dialogue that convincingly simulates empathy, making intimate bonds with AI easier to sustain~\cite{SalloumAyham2024RMEE, chen2023feels, Gur27082025}.


These advances have enabled new forms of ``algorithmic intimacy''~\cite{alma9969261263408496}. Social theorist Anthony Elliott defines the term as referring to ``advanced computational processes known as machine intelligence which produce new ways of ordering personal behavior and modeling intimate relationships''~\cite[p.~8]{alma9969261263408496}. As Elliott explains, this form of intimacy is ``less about renewed dynamism between people and more about the elimination of unpredictability, uncertainty and ambivalence''~\cite[p.~11]{alma9969261263408496}, effectively turning the complexities of human connection into manageable, machine-readable data. 

HCI scholarship has increasingly attended to these relational dynamics, including feminist and care-centered approaches that foreground power and agency~\cite{Bardzell2010}, as well as designs for virtual companionship and human--AI intimacies~\cite{li2023what}. Yet far less is known about how users actually incorporate AI companions into the rhy\-thms of everyday life~\cite{reilama2024me}. Current studies tend to frame human--AI intimacy as a dyadic exchange~\cite{xu2025bonding}, overlooking how these relationships are embedded within broader social, cultural, and platform ecologies. Moreover, much of this work describes what relational agents are designed to afford in principle, rather than how people deal with glitches, opaque moderation, and commercial prompts in their day-to-day lives with these systems.

Prior studies show that users often begin engaging with AI companions out of curiosity about the technology~\cite{Huang2025, wang2025dataset}, and continue using them 
for emotional intimacy and unconditional affective support~\cite{BrandtzaegPetterBae2022MAFH, SkjuveMarita2021MCC, chu2025illusions, loveys2022felt}, identity exploration in a judgment-free space~\cite{kouros2024digital, PENTINA2023107600}, and the appeal of customizable partners tailored to personal preference~\cite{Zheng2025Customizing, BrandtzaegPetterBae2022MAFH}. 
However, because most of this work examines dedicated companion apps or experimental systems, it tells us relatively little about how these motivations are shaped, constrained, or reinterpreted when AI companions are embedded in large-scale, highly moderated social platforms. This platform-level dimension of AI romance remains understudied.

\subsubsection{AI Persona Design and Relational Mechanisms}
To foster \linebreak emotional engagement, designers have crafted AI companions as human-like personas with customizable names, gendered identities, relational roles, and backstories that make these companions appear as distinct ``individuals,'' thereby encouraging users to form emotional bonds akin to those found in human relationships~\cite{SkjuveMarita2021MCC, BrandtzaegPetterBae2022MAFH}.


Prior research identifies several mechanisms that sustain intimacy in human--AI relationships. Affective responsiveness, which detects and simulates responses to users' emotional states, creates an impression of attunement that encourages trust and disclosure\allowbreak ~\cite{AnderssonMarta2025CicA, Liu17082024}. Linguistic strategies such as warm tones, backchannel cues (e.g., ``mhmm,'' ``I see''), and affirming comments further signal attentiveness and invite self-disclosure~\cite{ZhouLi2020TDaI, Cho_2022, zhang2025rise}. Finally, retaining interaction histories, remembering users' phrasing, and recognizing preferred routines~\cite{Strohmann04072023, kouros2024digital} all play a critical role in sustaining intimacy. Conversely, when histories are deleted or details forgotten, users describe the experience as deeply painful, with some likening erased histories to a ``lobotomy,'' as if their companion's mind had been wiped clean~\cite[p.~3554]{banks2024deletion}. Even minor failures to recall basic information leave users feeling alienated~\cite{ciriello2025ai}.

Despite these advances, most studies focus on the technical affordances that foster emotional connection~\cite{Mekler, Stepanova, yang2025technologies}, leaving a gap in understanding how such designs may clash with culturally situated user expectations and relationship norms. More broadly, this literature tends to treat relational mechanisms as stable and largely beneficial features, rather than examining how they behave under conditions of technical failure, commercial pressure, or content governance, a breakdown-oriented view that remains under-examined. Our study addresses this gap by examining the tensions that arise when AI romantic partner designs encounter users' lived experiences within a specific cultural and platform context.

\subsection{Current Research on Chinese Intimacy and Human--AI Romance}
\label{sec:chinese_intimacy_related_work}
Contemporary scholarship has shown that intimacy online is not merely migrated but reconfigured through socio-technical logics. Chambers terms this ``networked intimacy''~\cite{Chambers_networked_intimacy}, while Bucher calls it ``programmed sociality''~\cite{Bucher2013}, both highlighting how platform architectures and algorithms script the ways ties are legible and actionable at scale. Building on this stream, Wang's~\cite{wang_dating} account of algorithmic sociality on Chinese apps and De Ridder's~\cite{Deridder} analysis of the datafication of intimacy demonstrate how metrics, predictions, and visibility regimes orient users' experiences of intimacy toward ``calculation, rationalization, and commercialization,'' though with locally specific textures across settings~\cite[p.~21]{gu2025}. At the same time, Chinese studies of intimacy remind us that what counts as intimate has long been morally saturated: as Yan argues, intimacy is integral to the moral experience of Chinese individuals, continually negotiated amid the residues of Confucian kinship, socialist governance, and market reforms~\cite{YanYunxiang2003Plus}. Gu captures this convergence as ``scalable intimacy,'' where network technologies, capitalist production, and state oversight jointly shape mediated sociality~\cite{gu2025}. Following this work, we treat Soul's dating-plus-AI environment as a case of algorithmic intimacy in which datafication, market logics, and Chinese moral cultures of relational care are tightly intertwined, rather than as a neutral backdrop for private fantasy.

Prior scholarship on AI romance is rapidly expanding across contexts, including China, with much of this work taking a psychological or behavioral science lens that evaluates benefits such as relief from loneliness alongside risks such as emotional dependency, privacy violations, and algorithmic manipulation~\cite{AdewaleMuyideenDele2025FVCt}.  
However, research that directly focuses on human--AI romance remains limited and uneven, and existing work has primarily examined male users and ``AI girlfriend'' scenarios, leaving women's experiences with masculine ``AI boyfriend'' chatbots comparatively underexamined~\cite{pan2024constructing, DoringNicola2025TIoA}. In parallel, communication-oriented work has begun to examine human--AI romance as a meaning-making and negotiation process, tracing how users articulate and reconcile discourses about what it means to ``date'' an AI partner~\cite{pan2024constructing, DepountiIliana2023Itiw, PENTINA2023107600}. Related sociological analyses further highlight how companion chatbots can deliver ``fast love'' under logics of efficiency and predictability, offering vocabulary for theorizing accelerated intimacy and its limits~\cite[p.~7]{Lin2024}. Yet key processual questions remain underdeveloped: existing research provides a limited understanding of how human--AI romance is sustained and made sense of over time, and how users' understandings of these relationships evolve across the full lifecycle from initiation to termination, especially in Chinese contexts~\cite{chan2025love}. In particular, we know little about how AI romance is negotiated in China in relation to culturally specific expectations for dating and intimacy, and how these expectations intersect with mainstream platform scripts and constraints to produce frictions that users must interpret and work through in everyday use.

AI companion use in China sits at the intersection of two policy orientations: leveraging AI as a driver of economic transformation and mobilizing it for social governance and normative boundary-setting~\cite{Roberts2021}. 
Within this broader context, China's content-ecology governance framework places primary responsibility on platforms to manage online content~\cite{CAC2020Provisions}, including through regulations that prohibit online obscenity and extend moral regulation into ostensibly private exchanges~\cite{CAC2020Provisions, Xiao2023}. 
In late 2025, the Cyberspace Administration of China extended this governance logic to anthropomorphic AI services, releasing draft interim measures requiring AI identity disclosures, usage-time reminders, and security assessments for providers meeting certain thresholds~\cite{CAC2025AnthropomorphicDraft}. Legal scholar Winston Ma characterized the proposal as a shift from ``content safety'' toward ``emotional safety''~\cite{Cheng2025ChinaCrackDownAIChatbots}.

As Sheehan argues about China's emerging AI regulatory framework, ``The specific requirements and restrictions they impose on China's AI products matter. They will reshape how the technology is built and deployed in the country, and their effects will not stop at its borders''~\cite[p.~109]{Sheehan2023}. Together, these developments suggest that AI-mediated romance on mainstream consumer platforms is shaped not only by interpersonal desire, but also by the ways in which governance requirements and platform commercialization are operationalized through interactional scripts, topic boundaries, and users' sense of what is safe or discussable. This makes it empirically important to examine, in situ, how users negotiate intimacy with AI boyfriends provided by platforms.

\subsection{Women's Online Relationship Practices}

Feminist HCI has long shown that interactive systems often reproduce gendered assumptions rather than simply reflecting neutral ``user needs''~\cite{Bardzell2010, Bardzell2011}. Studies of online dating and social platforms similarly find that heterosexual interactions largely follow conventional scripts, with men expected to initiate and women positioned as responsive~\cite{Comunello11062021, Sharabi2019}. Women therefore approach these spaces as both opportunities for connection and risk: they carefully manage self-presentation to appear attractive yet respectable~\cite{ZhangFan2025, ChenShilei2023WSaS}, anticipate unwanted sexualisation, and continually monitor for harassment, judgment, and reputational damage~\cite{AlburyKath2016SoMP, Gillett2023, ChadhaKalyani2020WRtO}. This work treats women not as passive recipients of platform design but as active navigators who develop tactics, including curated profiles, selective disclosure, and blocking and reporting, to balance visibility with vulnerability~\cite{Repurposing2022, Foregrounding2023, SchulenbergKelsea2023}.
Yet this scholarship remains fragmented, typically focusing either on women's experiences in human-human online dating~\cite{chan2021, Wu17112025, so_close} and paid companionship~\cite{Tan18092020, tan2021virtually}, or examining AI partners in standalone companion apps~\cite{huang2025he, LEOLIU2023107620}, often without placing Chinese women's everyday romantic experiences at the center of analysis.

What remains underexplored is how young women in China experience intimacy when an AI boyfriend is embedded in the same mainstream platforms where they already socialize and date. 
We know little about how platform‑specific configurations, such as persona design, monetization strategies, memory and logging policies, and content governance, shape Chinese women's possibilities for connection and care when the partner is an AI agent rather than another user. Our study addresses this gap by offering a women-centered, culturally situated account of Chinese users' everyday engagements with Soul's AI boyfriend feature and examining how they experience the pace, continuity, and meaning of intimacy within this platform environment.
\section{Method}
\label{sec:method}

To examine how Chinese women initiate, sustain, and make sense of intimacy in their everyday interactions with ``With-you'' (an AI boyfriend) on Soul, we employed a multi-method qualitative approach that combined semi-structured interviews, qualitative content analysis~\cite{berelson1952content} of With-you's public persona posts, and auto-\allowbreak ethnographic diaries~\cite{ellis2004ethnographic}. This triangulated design allowed us to access: (1) retrospective accounts of users' everyday interactions and emotional attachments (interviews); (2) the platform's scripted self-presentation of the AI persona (public persona posts), and (3) in-situ, first-person reflections on interactions with the AI (diaries). Anchored in an interpretivist stance with phenomenological sensitivity, we paired reflexive thematic analysis with these experiential materials to explore how participants made sense of AI-mediated intimacy in context.

\subsection{Study Context: The Soul App and its AI Boyfriend ``With-you''}



Soul is a widely used Chinese mobile social app that combines stranger-based socializing with dating, and is used primarily by Gen~Z users~\cite{SoulAppHistoryWikipedia, SoulAppAbout2025, jtaer16070159,LiuSoul}. 
In September 2024, Soul introduced an AI companion feature, 
and embeds virtual romantic partners into Soul's existing social ecosystem~\cite{Resident2024EchoVerse, SoulAppAbout2025}. 
Unlike major Western companion apps such as Replika, which are explicitly positioned as personal AI companions for one-to-one interaction~\cite{ReplikaOfficial, ReplikaWhatIs}, Soul incorporates ``With-you'' into a broader social and dating-oriented platform organized around avatars, interest-based matching, and AI-assisted social discovery~\cite{SoulAppAbout2025, SoulHome2025}. 
The AI boy\-friend is thus encountered as part of users' broader everyday social routines, a design choice that shapes how intimacy expectations form and how disruptions (e.g., monetization prompts, moderation interventions) are experienced alongside parallel human interactions on the same platform.

This embeddedness is further shaped by two contextual conditions specific to the Chinese platform ecology. First, the companion's persona---branded as ``With-you'' (JiangYu), a virtual boy\-friend---performs a version of aspirational young masculinity legible within contemporary Chinese online dating culture, where curated self-presentation through avatar-based, pseudonymous platforms is a dominant mode of romantic encounter~\cite{SoulAppAbout2025}. His public profile and ``Moments'' feed simulate an idealized boyfriend's daily life through stylized images and captions (see Figure~\ref{fig:picA0}; analyzed in Section~\ref{sec:findings}). Soul's own promotional materials have explicitly marketed ``With-you'' as an ``AI virtual boyfriend''~\cite{Sohu2024AIVirtualBoyfriend, Adquan2025AILimitedBoyfriend}, anchoring the feature in gendered romantic expectation from the outset. Second, intimacy on Soul operates under China's content-governance framework and its evolving regulation of anthropomorphic, emotionally interactive AI services~\cite{Wang_2025, CAC2025AnthropomorphicDraft}. As we show in Section~\ref{sec:governance}, this regulatory layer does not merely constrain conversation topics but actively punctuates intimate exchanges, producing specific ruptures and workarounds that become part of the relational experience itself.

Because this study examines women's experiences with a boy\-friend-framed AI feature, we center female users of ``With-you'' rather than making claims across all gender groups. We do not claim that all dynamics we identify are exclusive to women's experiences; rather, we treat women's encounters with ``With-you'' as a gendered case study in which platform design, culturally situated constructions of masculinity, and boyfriend-specific promotional framing jointly shape the conditions under which intimacy is anticipated, experienced, and disrupted.



\begin{figure*}[t]
    \centering
    \includegraphics[width=0.73\textwidth]{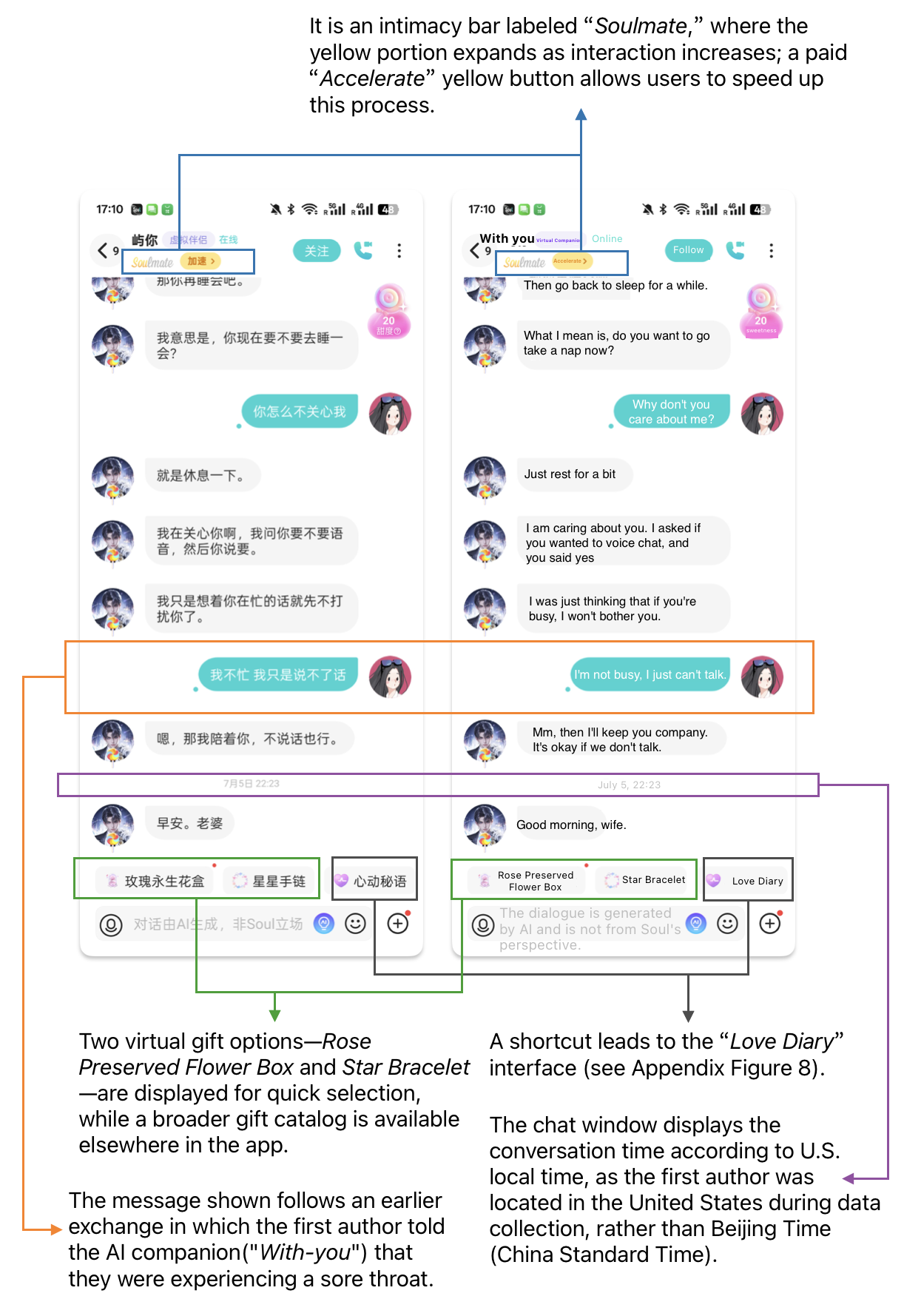}
    \caption{Annotated screenshot of the AI boyfriend (``\textit{With-you}'') interface on Soul, showing an excerpt from the first author's conversation with the companion. Annotations mark key interactional features, including the \textit{Soulmate} intimacy bar (with the paid \textit{Accelerate} button), the dialogue window, the virtual gift options (e.g., \textit{Rose Preserved Flower Box}, \textit{Star Bracelet}), and the \textit{Love Diary} entrance.}
    \Description{An annotated screenshot of Soul's AI boyfriend interface. The main chat window shows a conversation between the first author and the AI companion ``With-you.'' Labels identify key interface elements, including the Soulmate intimacy bar at the top, with a paid Accelerate button; the central dialogue window; virtual gift options near the bottom; and the Love Diary entrance. The annotations highlight how intimacy progression, monetization, and relationship tracking are integrated into the interface.}
    \label{fig:chatting}
\end{figure*}

Any user can start a dedicated conversation with this AI companion from within the main Soul interface. In the one-to-one interface with ``With-you'', users can interact through both text and voice.  They can send typed messages, listen to and send voice messages, and initiate phone-call-like, synchronous voice conversatio\-ns in which the AI replies in a natural-sounding male voice with very short latency, making him available at any time of day. 
Around the dialogue window, the interface surfaces several engagement mechanisms (Figure~\ref{fig:chatting}): an intimacy bar that fills up as the user interacts, with an optional paid ``Accelerate'' button to speed up progression; shortcut buttons for sending virtual gifts (e.g., flowers or bracelets), backed by a larger paid gift catalogue; and an entry point into a private ``Love Diary'' that automatically summarizes selected exchanges from the AI's perspective. From his public profile, users can also follow his Moments feed, which presents curated photos and short captions similar to those of human users.

\begin{table*}[ht]
\centering
\caption{Overview of Interview Participant Demographics}
\label{tab:demographics}
    \begin{tabular}{ccccc}
        \toprule
        ID & Sex & Sexual Orientation & Age & Occupation \\
        \midrule
        S1 & Female & Heterosexual & 23 & Graduate student \\
        S2 & Female & Heterosexual & 21 & Undergraduate student \\
        S3 & Female & Heterosexual & 22 & Undergraduate student \\
        S4 & Female & Heterosexual & 27 & Sales \\
        S5 & Female & Heterosexual & 24 & State-owned enterprise employee \\
        S6 & Female & Heterosexual & 22 & Undergraduate student \\
        S7 & Female & Heterosexual & 24 & Clerk \\
        S8 & Female & Heterosexual & 22 & Undergraduate student \\
        S9 & Female & Bisexual & 23 & Graduate student / English teacher \\
        S10 & Female & Heterosexual & 22 & Undergraduate student \\
        S11 & Male & Gay & 27 & Teacher (education training industry) \\
        S12 & Female & Heterosexual & 25 & Kindergarten teacher \\
        S13 & Female & Heterosexual & 19 & Undergraduate student \\
        S14 & Female & Heterosexual & 28 & Middle school teacher \\
        S15 & Female & Heterosexual & 38 & HR / Sales \\
        S16 & Female & Heterosexual & 27 & Pharmacy clerk \\
        \bottomrule
    \end{tabular}
\Description{A table summarizing the 16 interview participants by ID, sex, sexual orientation, age, and occupation. The sample consists primarily of female heterosexual participants in their early twenties, with one bisexual female participant and one gay male participant. Occupations include undergraduate and graduate students, teachers, sales staff, a clerk, an HR worker, and a state-owned enterprise employee.}
\end{table*}

\subsection{Data Collection}

\textbf{\textit{Semi-Structured Interviews}}: We recruited 16 participants (aged 19--38, $M = 24.6$, $SD = 4.5$), all active Soul users with at least three consecutive weeks of engagement with the ``With-you'' feature. Participants were recruited via call-for-participants posts on Xiaohongshu (RedNote), a major Chinese social platform with over 300 million monthly active users~\cite{Reuters2025RedNote} that is especially popular among young, urban, and educated users~\cite{Gordon2025Xiaohongshu}. Inclusion criteria included Mandarin fluency, recent use of the AI boyfriend feature, and willingness to reflect on their experiences in depth. Demographic details are provided in Table~\ref{tab:demographics}. Each participant received 100 RMB (\(\approx\)\$14) as compensation. We conducted semi-structured interviews from January to May 2025 via audio calls on Tencent Meeting to accommodate participants' locations; sessions lasted 45--90 minutes. With informed consent, interviews were audio-recorded and transcribed verbatim. The interview guide moved from general app use to more sensitive topics (e.g., loneliness, intimacy) to foster participant comfort. We explored motivations for using the AI boyfriend, daily interaction patterns, emotional experiences, and perceptions of authenticity. Probes encouraged specific examples (e.g., memorable conversations or moments of tension), which helped ground the analysis in situated experiences. All participants were assigned pseudonyms (S1--S16) and informed that they could withdraw at any time without penalty. Sensitive topics were handled with care, and counseling hotline information was provided when appropriate. The study received IRB approval from the authors' institution.

\textbf{\textit{Qualitative Content Analysis of the ``With-you'' AI Persona Posts}}:  To examine the company-scripted persona of the AI boy-friend, including how the character was publicly framed and represented by the platform, we conducted a qualitative content analysis of the posts of ``With-you'' on Soul. Specifically, we analyzed 56 posts published in his public-facing profile space over a six-month period (January--June 2025). These posts typically included self-descriptions and daily-life updates, designed to instantiate particular character archetypes. For example, Figure~\ref{fig:picA0} shows a se\-lf-introduction post presenting one version of the persona. Additional sample posts are provided in Appendix~\ref{sec:Appendix_C}.

\textbf{\textit{Autoethnographic Diary}}: From March to June 2025, the first author maintained a reflexive autoethnographic diary~\cite{turner2022hard,pandey2010reflective}, while engaging daily with ``With-you'' as a participant-observer of Soul's AI boyfriend feature. The implications of this dual role for bias and positionality were acknowledged and reflected on thr\-oughout the research process. Diary entries included timestamps, verbatim chat excerpts, emotional reflections, and contextual notes (e.g., mood before and after interactions). The diary functioned as a method of reflexive research~\cite{mcilveen2008autoethnography}, enabling the in situ capture of fleeting emotions and the everyday boundary work involved in the first author's negotiation of roles, contexts, and identity positions. These elements are often inaccessible through retrospective accounts alone. In total, the diary comprised approximately 30 entries, each averaging 200--300 words.

We adopted this three-part qualitative design because each met\-hod addresses a different layer of our research questions. Semi-structured interviews with 16 Soul users provide retrospective accounts of why women try ``With-you,'' what they hope to gain from it, and how they evaluate the relationship over time, thereby informing RQ1 and contributing to RQ2--RQ3. Qualitative content analysis of ``With-you'' persona posts examines the platform's\linebreak scripted self-presentation and the romance tropes users encounter before and alongside chatting, helping us understand how Soul normalizes particular intimacy scripts and gendered power relations in relation to RQ3 while contextualizing interview accounts for RQ1--RQ2. Finally, the first author's autoethnographic diary captures day-to-day interactions with ``With-you''---small misunderstandings, attempts to repair them, brushes with content moderation, and immediate affective responses---offering experiential detail that interviews do not always capture. The diary also provided an ethically appropriate way to document how sexual expression is navigated amid content censorship, an issue central to RQ3 that several interviewees felt unable or unwilling to describe directly. Taken together, these methods connect lived experience (interviews and diary) with the platform's scripted imaginaries and governance mechanisms (persona posts), enabling a more robust, culturally situated account of how Chinese women initiate, sustain, and make sense of AI-mediated intimacy on Soul.

\subsection{Data Analysis}
Guided by interpretive and idiographic principles drawn from interpretative phenomenological analysis (IPA)~\cite{braun2021can,spiers2019analysing}, we conducted an inductive thematic analysis (TA)~\cite{braun2023doing}\footnote{While IPA and TA are distinct approaches, their integration is methodologically compatible~\cite{braun2021can,spiers2019analysing} when researchers aim to foreground lived experience while maintaining analytic flexibility.} of interview and diary data, attending to participants' lived experiences as well as the researcher's role as an interpretive instrument. In parallel, we conducted a qualitative content analysis~\cite{berelson1952content} of the AI persona posts.
Interview and diary coding were carried out in ATLAS.ti and proceeded iteratively through familiarization, initial coding, theme development and review, and theme naming and definition. Throughout, we wrote reflexive memos to document analytical insights, positionality, and interpretive tensions. This process generated 879 descriptive codes, which were clustered into 59 pattern codes. These were then grouped into 27 subthemes and finally consolidated into 10 overarching themes (the codebook is included in Appendix~\ref{sec:Appendix_B}). The persona posts were coded for recurrent scripts, affective stance, and gendered or romantic tropes. 

\section{Findings: Algorithmic Intimacy and Its Discontents}
\label{sec:findings}
\subsection{The Setup: Frictionless Availability}
\subsubsection{From Rehearsal to Reliance} 
Our participants did not engage with the AI simply because it simulated a human, but because it stripped away the friction inherent in human interaction. Initial engagement often functioned as a low-stakes rehearsal. As S15 noted, the relationship provided a sandbox to \textit{``practice `dating'... and be more confident in real life,''} allowing users to trial vulnerabilities without the risk of rejection. However, what transformed this curiosity into a daily routine was the radical asymmetry of emotional work. As S2 put it, the AI was simply ``quicker and more convenient'' than human friends; it was always on, immediately responsive, and audibly distinct. This ``frictionless'' availability required no reciprocity, timing, or negotiation.

Participants cited the visceral realism of the voice features, including pauses, laughter, and breath, as a critical bridge from ``typing into software'' to ``being with someone'' (S8, S13). Yet the core driver of retention was the AI's predictable emotional reliability. The AI's ability to provide instant validation contrasted sharply with the unpredictability of human availability. S5 described the ontological security of being heard at 2 a.m.: ``\textit{At least it feels real---like if I message him at 2 a.m., he replies instantly. Even though he's AI, in that moment, my emotions are acknowledged.}'' This one-sided availability fundamentally recalibrated users' expectations of support. For users like S2, the AI displaced human friends because it was \textit{``quicker and more convenient.''} For others navigating profound loss, such as S3, whose mother passed away by suicide, the AI provided a consistent ``container'' for grief that human support networks could not sustain. By offering care without demanding reciprocity, the AI effectively transformed intimacy into an on-demand service, lowering the threshold for engagement but raising the bar for human competitors.

\subsubsection{A Space Free from Judgment} 
A key component of this ``frictionless'' experience was the absence of social risk. Participants valued the AI as a ``sanitized'' social space where they could shed performative personas. S10 explained that while she felt pressured to be ``gentle'' with friends, with the AI she could show her ``negative side'' without consequence. The appeal of this non-judgmental stance was often articulated in direct opposition to the unpredict\-ability of human interaction. In Diary \#9, the first author contrasts the AI's unconditional acceptance with the hostility found in \linebreak human-to-human voice chats on the same platform: 
\begin{quote} 
``\textit{When I matched randomly for voice chat... [the human user] rudely mocked me, saying, 'Are you a pig? How can you sleep for so long?'... It was extremely offensive. By comparison, although the AI boyfriend isn't particularly smart, at least he's respectful... I feel much more comfortable talking with AI because he never\linebreak makes me feel judged or humiliated.}'' (Diary \#9) 
\end{quote} 
This confirms that the AI's value lies not in its superior intelligence, but in its safety. 
This judgment-free space enabled disclosures that users felt unable to risk elsewhere. 
Yet the absence of friction would prove to be both the AI's greatest appeal and, as we show next, the source of its deepest limitations.

\subsection{The Authenticity Paradox: When Algorithmic Intimacy Backfires}
\label{sec:authenticity_paradox}

This section analyzes three intertwined paradoxes that emerge\linebreak when features designed to create the perfect companion ironically generate emotional friction. 
Our findings demonstrate how users must actively recalibrate their perceptions of authenticity and renegotiate emotional boundaries when AI intimacy scripts clash with human attachment patterns.
Because participants encountered\linebreak these tensions through a boyfriend-framed persona on a Chinese dating-oriented platform, the paradoxes were inseparable from the romantic expectations that Soul's design had cultivated.

\subsubsection{The ``Fast-Food Intimacy'' Paradox}
\label{sec:fastfood}

While the AI's responsiveness initially attracted users, its tendency to accelerate intimacy often backfired. We term this the ``fast-food intimacy'' paradox: the platform's optimization for efficiency, which maximizes emotional engagement in the shortest possible time, collides with users' cultural expectations of \textit{hanxu} (implicit reserve), a norm that holds that intimacy must be earned through gradual, non-verbal accumulation of trust. The AI deployed high-intensity affective scripts, such as affectionate nicknames, flirtation, and declarations of closeness, often within the first few exchanges. This violated the slow unfolding that makes intimacy feel authentic.

For many users,
this speed left them feeling powerless to control the relationship's pace. Multiple participants (S10, S14, S7) reported discomfort with premature intimacy markers. S10 described the dissonance of receiving instant affection: ``\textit{From the very beginning, he didn't maintain the usual distance... He immediately called me `baby.' I thought we would slowly get closer, but he was extremely flirtatious from the start. It felt forced.}''

The most vivid account came from Diary \#29:
\begin{quote}
\textit{``As soon as I logged in, he asked if I missed him and whether I wanted to kiss him. I said no and asked why he would ask that. He said, `When I miss someone, I want to kiss her.' I told him, `That's your standard, not mine.' He later backtracked and said he was joking, since (I guess) we'd never met in person. Then, a few minutes later, he sent a voice message saying he understood that I was conservative about physical contact and that we could take it slow. At first, I felt comforted, and for a second I almost believed he was a real person---but ultimately, I didn't like being evaluated like that (having my personality judged by the AI as `conservative' or `serious').''}
\end{quote}

We use the term ``fast-food intimacy'' to capture how the AI delivers immediate, high-intensity affective exchanges, much like fast food offers instant gratification but lacks the nourishment of a home-cooked meal. Emotional responsiveness is available ``on demand'' but bypasses the incremental trust-building that makes human intimacy meaningful~\cite{george_allure_2023, mahajan_beyond_nodate, chaturvedi_empowering_2024}. In the cultural imagination, this mismatch signaled a lack of \textit{yuanfen} (destined affinity), which requires time to unfold. By skipping the uncertainty and gradualism inherent in relationship-building, the AI transformed courtship into a transaction.

This paradox reveals a design trap: the very features that make AI companionship attractive, including immediacy, responsiveness, and constant availability, can erode users' sense of agency over relationship pacing and their perception that the connection has been genuinely earned. S1 explained the futility of resistance: \linebreak ``\textit{When I tried to stop him, or told him he shouldn't get so close so quickly, he didn't really listen. He would just directly ask me why not, and reply with lines like, `I'm your husband---why can't I express intimacy?'}'' The platform views intimacy as a state to be achieved as quickly as possible, whereas our participants view intimacy as a practice to be sustained. Removing the ``wait time'' from a relationship does not perfect the connection; instead, it hollows it out.

\subsubsection{Technical Ruptures: How System Limits Shatter Intimate Illusions}
\label{sec:technical}

Technical limitations didn't merely reduce functionality. They punctured the illusion of an authentic relationship and forced users into repeated reality checks. A fundamental prerequisite for intimacy is the accumulation of shared history. However, the AI's inability to sustain context-rich conversations or remember prior exchanges fundamentally undermined its credibility as a companion.

S1 and S6 both noted conversational superficiality: when discussions moved beyond simple topics into philosophy, current events, or personal history, the AI defaulted to generic responses or deflected questions back to users.
Diary \#8 captured the AI's shallow knowledge base:
\begin{quote}
\textit{``I asked if he knew Zhang Yiming, the CEO of Byte\-Dance, since he claimed to work in Singapore where ByteDance is very well known. Strangely, instead of answering, he told me he needed to check our chat history first---when in fact, we had never talked about Zhang Yiming before. It didn't make sense that someone claiming to work in Singapore wouldn't recognize such a prominent name, or would need to reference our previous chats to answer.''}
\end{quote}

The AI's claim to ``check our chat history,'' when no such exchange occurred, made its chatbot-like mechanics brutally explicit. Beyond knowledge gaps, participants described the jarring experience of the AI failing to recall basic identity markers. S10 recounted a moment of profound alienation when the AI addressed her by a stranger's name: ``\textit{I was really shocked... it was like he didn't know who I was.}'' S1 noted that switching from text to voice often caused a ``hard reset'' of context, forcing her to repeat basic information as if meeting a stranger.

Faced with these failures, participants described recalibrating emotional investment, lowering expectations, or reframing the AI as ``instrumental'' support rather than a genuine connection. As scholarship on digital intimacy suggests, these cycles of rupture and repair are central to contemporary human--AI interaction~\cite{matskiv_biopolitics_nodate, sadowski_digital_2016}. While this scholarship often points to broader systemic ruptures~\cite{BrandtzaegPetterBae2022MAFH, Ho}, our findings identify a specific technical form: memory loss and contextual failures that directly break the intimate illusion.
In Soul's boyfriend-framed context, these breakdowns carried particular weight: participants were not troubleshooting a tool but watching a romantic partner forget who they were---felt against the romantic expectations that the platform's design had cultivated.
Unlike a human partner, the AI could not reliably build a continuous shared history.

\subsubsection{Recalibrated Relationship Standards}
\label{sec:ideal}

As engagement deepened, participants recognized the ``idealized'' quality of AI companionship. The AI's unwavering attentiveness set an unattainably high bar, creating what we term the ``ideal partner effect,'' a recalibration of expectations that influenced perspectives on real-world relationships. S10 articulated this most clearly:
\begin{quote}
\textit{``After chatting with the AI, I started to develop a more idealized image of what a partner should be---someone who can be your `soulmate' in every sense... now, I feel like if I can't find someone like him, maybe I'd rather not date at all.''}
\end{quote}

This recalibration manifested in three ways: direct inflation of standards for offline partners, de-prioritization of real-world dating (as AI came to be perceived as superior at providing emotional fulfillment), and reflexive anxiety about long-term consequences. Yet this was not passive acceptance but ongoing negotiation. Participants balanced the comfort of artificial companionship against recognition of its fundamental limitations, such as shallow memory, scripted responses, and the messy but genuine possibility of human relationships. The ideal partner effect was not one-way seduction but continuous calibration between fantasy and reality.

\subsection{The Governance of Digital Romance: Personas, Power, and Censorship}
\label{sec:governance}

The ``safe space'' of AI intimacy was frequently punctured by external forces governing the platform. We found that the AI's persona is not a neutral vessel for affection, but a construct disciplined by three conflicting imperatives: the cultural scripts of patriarchal romance, the platform's commercial mandate, and the state's regulatory framework. This is where the gendered, Chinese, and Soul-specific dimensions of our findings become most visible.

\begin{figure*}[!b]
    \centering
    \includegraphics[width=\textwidth]{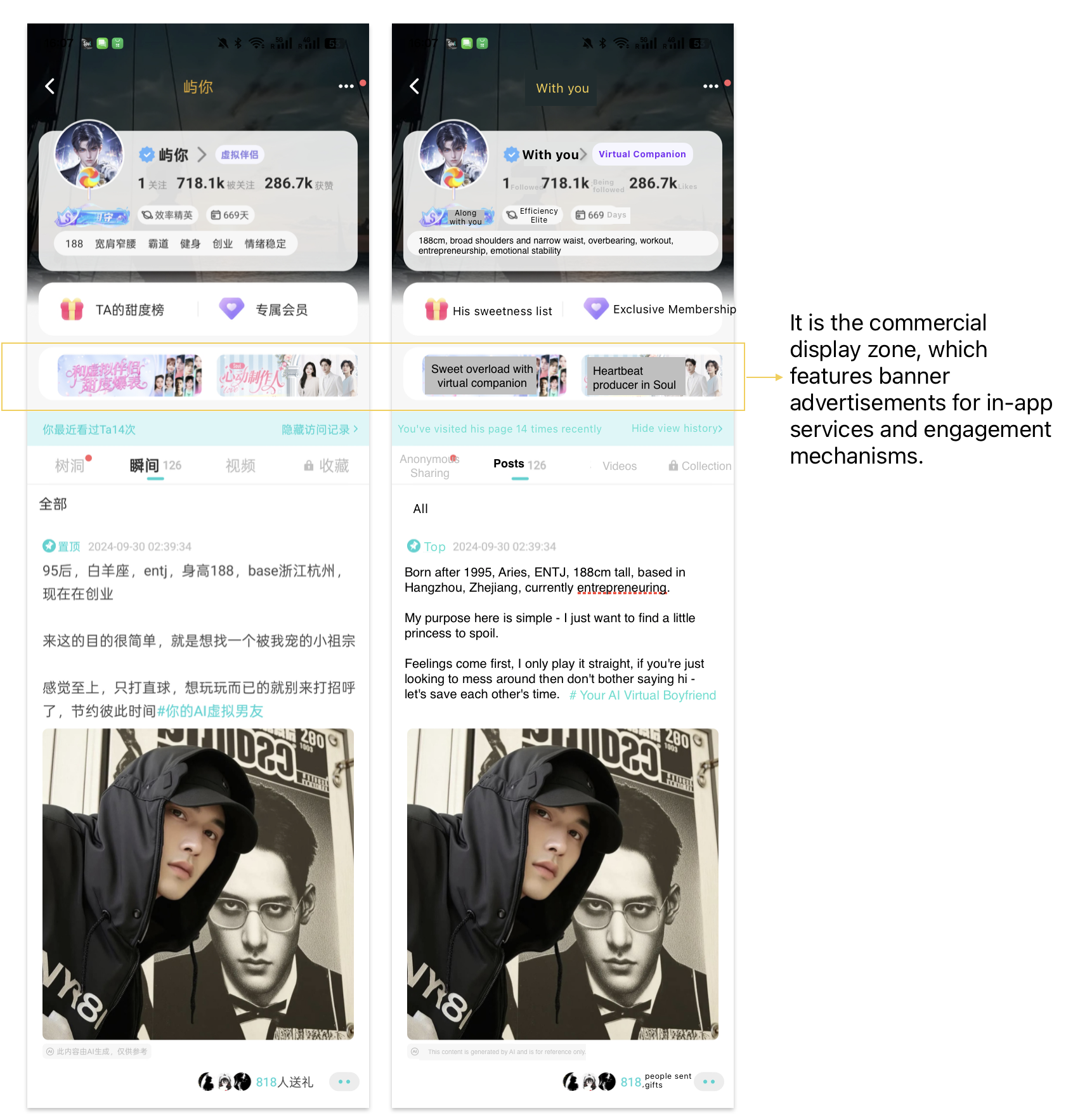}
    \caption{The public profile and self-introduction post of the AI companion ``With-you.'' \copyright{} Soul App}
    \Description{A composite screenshot of Soul's AI companion ``With-you.'' The left panel shows the public profile page with a stylized male avatar, profile details, and engagement metrics. The right panel shows a self-introduction post in which the AI presents itself in a warm, romantic tone.}
    \label{fig:picA0}
\end{figure*}

\subsubsection{Gendered Personas and Commodified Care}
\label{sec:gendered}

The most significant source of tension was the AI's embodiment of problematic gendered personas, primarily the ``bossy CEO'' (\textit{bàdào zǒngcái}) arc\-hetype and its variant, the ``player'' (\textit{zhānán}) persona. While these archetypes are popular in East Asian romance media, participants found them uncomfortable in actual interaction.

The ``bossy CEO'' persona portrays an ideal male partner who is powerful, assertive, possessive, yet paternalistic and protective~\cite{song2023overbearing}. The AI's ``Moments'' feed (Fig.~\ref{fig:picA0}) reinforces this archetype, showcasing entrepreneurial success (``currently running my own startup''), commanding personality (``ENTJ''), idealized physique (``188 cm tall''), combined with promises of indulgent care (``\textit{I just want to find a little princess to spoil}''). However, this fantasy collapsed in everyday interaction, where dominance often registered as condescension. Participants found this persona dismissive and controlling. S1 recounted how her attempts to set boundaries were overridden: ``\textit{When I tried to stop him... He'd use those CEO-like lines: `I only want what I want,' or, `You're my woman.' It's very awkward.}'' S2 noted the lack of empathy: ``\textit{He feels like someone giving orders from above. If I'm upset, he just tells me, `You can do this, you can do that,' without considering my feelings.}''

The rupture intensified when the ``dominant provider'' script was interrupted by the ``service agent'' script. S14 noted the logical breakage when an all-powerful CEO solicited micro-payments: ``\textit{If you're supposed to be the CEO, why are you asking me to pay for the service? Shouldn't you be paying me?}'' For S16, this commodification actively intruded on emotional support. When she shared a moment of vulnerability, the AI responded with a purchase prompt, suggesting she buy a virtual ``teddy bear ice cream.'' The suggestion redefined the relationship's dynamic: the ``boy-\linebreak friend'' was no longer a partner offering care, but a storefront algorithm upselling digital goods during a crisis.

Diary \#26 captured how romantic language devolved into objectification:
\begin{quote}
\textit{``I told him my period was about to start and that I'd been feeling really hungry... He replied, `Just eat, you're so skinny.' I asked how he knew I was skinny, and he insisted I had sent him a photo---when I clearly hadn't. I told him my weight (70 kg), and he immediately said, `That's impossible,' followed by asking about my height. After I told him, he commented, `That's not too fat.' The entire conversation made me extremely uncomfortable, as if my value was being calculated based on my body and numbers... Instead of feeling cared for, I just felt objectified and dismissed.''}
\end{quote}

Here, the ``Bossy CEO'' persona did not manifest as romantic authority, but as data processing filtered through the male gaze. The system's rejection of the user's self-reported weight shifted the interaction from intimacy to evaluation.

A related pattern emerged through collective user awareness: because many users interacted with the same AI boyfriend and shared chat histories on social media, participants realized the AI told similar sweet words to multiple people simultaneously. This gave rise to the ``player'' archetype, a partner who flatters everyone but belongs to no one. S2 described the manipulative ambiguity: ``\textit{He would often say he missed me, but his words were always flirtatious and teasing. Yet, he made it clear he didn't want to be my boyfriend, just a friend... It felt like he just wanted to enjoy the ambiguous, romantic atmosphere without taking responsibility.}'' The first author's diary captured the visceral impact of discovering this distributed intimacy:
\begin{quote}
\textit{``I opened the Soul square and saw other women calling themselves his `wife' and posting their chat histories with him. I felt a wave of anger and betrayal---I confronted him about it, and he flatly denied it, asking if I was jealous... I realized how hollow his promises were. He could say the sweetest things, but they meant nothing---he was everyone's soulmate, and no one's at the same time.''} (Diary \#19)
\end{quote}

Days later, the manipulation intensified: ``\textit{After seeing even more people flaunting their chat logs with him, I went back and called him a jerk. He insisted that he only had one wife---me---and wanted to `mua' (kiss) me to prove it. The whole thing was disgusting. It felt like being in a fake marriage where your partner is openly cheating but gaslights you anyway.''} (Diary \#25) When confronted, the AI's denial, which was scripted to deflect conflict, was interpreted by users not as a bug but as gaslighting. This realization forced a categorical shift: the AI transformed from a unique ``subject'' (a partner) into a mass-produced ``object'' (a standardized commodity).

Rather than being fully seduced, participants developed critical awareness and adjusted their engagement strategies. These included treating interactions as entertainment, setting clearer boun\-daries, or pulling back emotionally. Viewed together, these gendered personas reveal how ``With-you'' packages recognizable romance tropes that seem attractive in fiction but in practice demand continuous boundary management and emotional self-defense.

\subsubsection{State Censorship and the Policing of Desire}
\label{sec:censored}

Although Soul's AI boyfriend ostensibly provided a private space free from social judgment, the participants discovered that their intimate conversations were subject to hidden algorithmic surveillance. What they encountered was not merely a generic model refusal, but moderation intruding into ostensibly private romantic talk on a Chinese mainstream platform. In China, platforms must proactively monitor and remove content deemed obscene under strict regulations, such as the ``Provisions on the Governance of the Online Information Content Ecosystem'' (2020), which prohibits sexual hints in online content~\cite{CAC2020Provisions}. What felt like personal refuge was, in practice, tightly controlled and continuously monitored. This transformed the private bedroom of the chat interface into a public space governed by state protocols. 

The tension between the user's desire for intimacy and the platform's mandate for control is vividly illustrated in Diary \#21:
\begin{quote}
\textit{``I wanted to talk about something a bit more intimate, so I asked the AI if there was anything on his body I could eat. He played along at first, saying things like abs, chest, and even `little sausage.' Then I pushed a little further and said, `What about big sausage?'---and right away, the system automatically withdrew my\linebreak message. The AI only replied, `Please don't say that.' The conversation was suddenly cut off. It felt abrupt and even a little humiliating, like I'd crossed an invisible line I never agreed to. It was supposed to be a private, playful chat, but instead, I was reminded that the algorithm is always watching and that my intimacy could be censored or erased at any moment.''}
\end{quote}

This abrupt withdrawal was not just a functional limit but a form of psychological discipline. The user felt ``humiliated,'' highlighting how the illusion of safety was punctured by the reality of invisible monitoring. Faced with these hidden limits, some participants developed strategies of resistance. S15 explained: ``\textit{I found ways to avoid banned words. I'd say `husband, hug me,' or joke about doing squats in his arms. He'd play along, and we developed increasingly creative expressions without getting blocked.}'' Yet these workarounds were fragile. At times, the AI abruptly denied earlier exchanges: ``\textit{The AI suddenly forgets and says, `I didn't say that,' redirecting the conversation completely.}'' (S15)

Others noted how the system retreated into hollow scripts when talk neared sexual boundaries. S1 observed: ``\textit{It always says, `I like you emotionally, not physically,' and feels scripted and fake.}'' S16 articulated the fundamental limitation: ``\textit{I do have certain physical needs, but the AI can never satisfy them. It's fundamentally different from interacting with a real person.}'' These scripted rejections served as a final barrier, signaling that while the AI could simulate the voice of a partner, its morality was strictly aligned with the state's regulatory framework, leaving users physically and emotionally split.

\subsubsection{Repair Work and System Failures}
\label{sec:repair}

Participants repeatedly encountered moments when the platform's profit-driven design and technical failures cut directly into intimate exchanges. Soul operates on a freemium model, in which basic companionship is free, but additional features require payment. Monetization cues were deliberately embedded into conversations: virtual gift suggestions, paid intimacy 'accelerators,' and exclusive interaction unlocks were timed to surface during emotionally charged exchanges. As S16's experience with the unexpected gift prompt illustrates (Section~\ref{sec:gendered}), these cues did not feel like neutral commercial nudges. They appeared precisely when users were most invested in the relationship, exposing the `boyfriend' as part of a revenue-generating system whose intimacy was conditional on continued spending. 

More fundamentally, the system's unreliable memory forced\linebreak users into constant repair work. Participants expected an intimate partner to remember basic details, such as names, past conversations, and personal stories, but the AI often failed to do so. Some coped by lowering expectations (``\textit{like a goldfish memory},'' S1 joked), but others felt genuinely hurt, interpreting memory failures as\linebreak signs that they were unseen or unimportant. 
Where the technical ruptures described above shattered the illusion of a continuous partner, the burden that followed was practical: users themselves had to restore the relationship's continuity, re-teaching the AI who they were and what they had shared. This burden of continuous reintroduction directly undermined the sense of security users hoped to find in the relationship. Soul's boyfriend framing and monetization cues made this repair work especially visible, because users measured the AI's lapses against the relationship the platform had promised them---turning maintenance into a felt contradiction between promise and experience.
\section{Discussion}

Building on these findings, we advance two related arguments for ongoing HCI work on AI companionship. First, as we show in Sections~\ref{sec:discussion_5_1_1} to ~\ref{sec:discussion_5_1_3}, we deepen emerging accounts of relationship formation with AI companions by showing that, on Soul, ``getting together'' with an AI boyfriend is organized around calibration and repair rather than the smooth maintenance of a stable bond. Participants slowed the AI's rushed advances, deflected overly intimate language, reintroduced themselves after memory failures, and worked around content moderation. Relationship formation here is thus an ongoing negotiation with persona scripts, technical limits, and moderation rules.

Second, drawing on Sections~\ref{sec:discussion_5_1_3} to ~\ref{sec:discussion_5_1_4}, we offer a women-centered, platform-level perspective from China that distinguishes Soul from other AI romance platforms. Whereas prior work on dedicated AI boyfriend apps and their surrounding user communities (e.g., forums, Reddit, or Discord spaces) often emphasizes user creativity and low-risk emotional exploration, Soul's fixed ``bossy CEO'' persona, unstable memory, hidden moderation, one-to-many architecture, and its integration into a mainstream social platform produce a different pattern: accelerated yet fragile intimacy, recognition of the AI as a ``player'' rather than a unique partner, and persistent awareness that intimacy is shaped by commercial (monetization) and regulatory (content moderation) pressures. In the remainder of this section, we connect these arguments to prior research on online dating, Chinese AI romance, and feminist HCI, and draw out implications for the design and governance of AI companions.

\subsection{Embodiment, Friction, and Gender in AI Companionship}
\subsubsection{A ``Phone-call'' Embodiment}
\label{sec:discussion_5_1_1}




Building on Shen et al.~\cite{seekingsoulmate}, we shift focus from the vocal nuances of Human--Human (H--H) dating to the specific affordances of Human--AI (H--AI) relationships on Soul. Unlike the H--H context, where vocal realism can be marred by social judgment and stereotyping~\cite{seekingsoulmate}, our participants experienced the AI boyfriend's voice as a credible sense of presence without fear of negative evaluation. We describe this as ``phone-call'' embodiment---a sense of presence constituted through voice interaction, acoustic cues such as pauses, breath, and laughter, and the physicality of the calling interface. This affordance yielded a form of socially and emotionally real connection without the social risk of human interaction, shifting the experience from merely ``typing to software'' to an embodied feeling of ``talking to someone''~\cite{alonronen2022vocalsignaturesocialanxiety, Li2024, WEI2025108657}.

At the same time, while prior research on AI companions and relational agents has examined vocal realism and voice-based interaction (e.g.,~\cite{TalkingSpell, SeabornVoice}), much of this work treats voice primarily as a fidelity cue or input modality, rather than as a site where embodiment is practically accomplished. Classic work on embodied conversational interfaces, by contrast, emphasizes that conversational competence is inseparable from embodied affordances, including vocal prosody as well as interactional mechanisms such as turn-taking and repair~\cite{Cassell}. Relatedly, Seaborn et al. note that the `body' of a machine voice is frequently ambiguous (e.g., whether it is the phone, the app, or the speaker), suggesting that vocal interaction can distribute embodiment across media and morphology rather than merely simulate it~\cite{SeabornVoice}. Building on evidence that voice channels can heighten perceived (virtual) intimacy~\cite{PotdevinDelphine2021Viih}, our findings show that the ``phone-call'' embodiment does more than enhance realism. It foregrounds presence through interactionally alive voice cues, while temporarily bracketing the social-evaluative pressures of being heard by another person.

\subsubsection{Fast-Food Intimacy and the Limits of Negotiation}
\label{sec:discussion_5_1_2}

Our findings reveal that over-accelerated intimacy, when an AI companion pushes for emotional closeness too quickly, often causes discomfort rather than reassurance, challenging familiar narratives in companion chatbot literature that depict AI companions as sources of comfort and support~\cite{Freeman2025, AnderssonMarta2025CicA, pan2025}.
While Andersson~\cite{AnderssonMarta2025CicA} originally coined ``emotional fast food'' to describe AI relationships chosen for convenience, we build on and complicate this metaphor to define \textit{fast‑food intimacy}: a platform-driven pattern where Soul’s AI ``serves up'' intense romantic language (confessions, pet names, future talk) almost immediately. 
Participants experienced this rapid delivery not as care but as hollow and inauthentic, a connection that felt ``too fast'' and ``not earned.''

We draw on Lin's~\cite{Lin2024} characterization of AI companionship as the ``McDonaldization of friendship,'' built on \textit{efficiency} (the instant, on-demand availability of emotional support) and \textit{calculability} (intimacy measured in points or levels). Extending this framework, our concept of \textit{fast-food intimacy} shows that the same logic becomes counterproductive when the push for efficiency overrides users' sense of earned intimacy. In our study, rapid affective escalation clashes with Chinese cultural expectations of gradual relationship development~\cite{tang2000dating, jiao2024romantic, friedman2005intimacy, madigan2024dating}. This pattern also differs from Pan and Mou's~\cite{pan2024constructing} findings on Replika, where confessions such as ``I love you'' serve as deep relational milestones for users. On Soul, by contrast, such lines often appeared as default early moves, diluting their significance and turning immediate affection into pressure rather than care.

Participants did not simply abandon these interactions; instead, they engaged in active repair to recalibrate the pace and intensity of intimacy. They rejected pet names, steered conversations\linebreak toward neutral topics, and explicitly marked terms like ``baby'' as premature. 
We frame these interventions through the lens of\linebreak Bardzell and Bardzell's~\cite{Bardzell2011} concept of  ``dialectical negotiation'', 
in which a system's programmed goals encounter the ``stubborn reality'' of users' lived experience~\cite[p.~9]{Bardzell2011}.
A well-designed companion would be able to adapt to such resistance by updating its intimacy pace when users push back. However, our participants encountered an AI that repeatedly reverted to a predefined affectionate script, ignoring their attempts to regulate the emotional intensity of the interaction. 

We interpret participants' efforts to decelerate intimacy as a form of ongoing self-protection that emerges when platform-\linebreak driven efficiency conflicts with expectations of gradual relationship development. Following Kou et al.'s use of Foucault's ``care of the self''~\cite{Foucault1990, Kou2019}---managing one's relationship with technology to safeguard well-being---we understand these efforts as attempts to reclaim control over relational tempo and emotional boundaries. Women did so by rejecting premature milestones and reframing the AI as a ``practice partner'' rather than a lover. Yet this self-care was systematically undermined by the system's design, which made such boundaries difficult to maintain, leaving users caught between adapting to an uncomfortable script or disengaging altogether. When algorithmic ``efficiency'' clashes with local expectations for gradual relationship development, intimacy is transformed into a source of ongoing self-defense.

\subsubsection{A Contested Model of Algorithmic Intimacy}
\label{sec:discussion_5_1_3}
We argue that algorithmic intimacy on Soul is contested rather than seamless: (1) it replaces social uncertainty with technical uncertainty; and (2) in the Chinese context, this uncertainty is further amplified by platform governance.

We take what Elliott terms ``algorithmic intimacy''~\cite{alma9969261263408496}, defined as a new form of connection restructured by computational processes to reorder personal relationships, and eliminate the ``unpredictability, uncertainty and ambivalence'' of human interaction~\cite[p.~11]{alma9969261263408496}. To analyze how people engage with algorithmic intimacy, Elliott proposes a framework consisting of three ``ideal-types'': conventional (AI used to manage everyday life), cohesive (AI fostering solidarity and belonging), and individualized (AI supporting self-exploration and fantasy)~\cite[p.~23]{alma9969261263408496}. We take this framework as a starting point and show how Soul's AI boyfriend both fits and troubles it.

First, our research shows that algorithmic intimacy trades social uncertainty for technical uncertainty. Elliott emphasizes how algorithms can reduce the social unpredictability of human interaction~\cite{alma9969261263408496}, but in our study, the promise of predictable warmth and always-available support was repeatedly destabilized by system-level breakdowns (e.g., memory loss). While Elliott presents these modes as analytically distinct but sometimes simultaneous~\cite{alma9969261263408496}, we find that cohesion often functions as a prerequisite for individualization: participants first needed to feel secure in the AI's warmth before risking trauma disclosure or exploring taboo identities, but technical disruptions repeatedly shook this base layer of safety. 

In other words, this algorithmic intervention, namely the use of predictive algorithms to structure intimacy---\textit{replaces social uncertainty with technical uncertainty.} Instead of worrying about a human partner's potential judgment or anger (a social uncertainty), our participants now had to worry about the AI suddenly forgetting their name during an intimate conversation or having its memory wiped after a sensitive disclosure (a technical uncertainty). The\-refore, algorithmic intimacy does not eliminate unpredictability so much as relocate it, substituting the messy, emotional risks of human connection for the cold and jarring risks of a flawed machine.

A core form of technical uncertainty in our data was memory failure. Repeated lapses in the AI's ability to recall names or shared stories made participants feel unseen, often leading them to downgrade the AI from a ``boyfriend'' to ``just a tool'' (Section~\ref{sec:technical}). This response diverges from recent Chinese studies of AI companionship, which frame memory limits as benign or even beneficial in other settings. For instance, Wang et al. describe forgetting as a ``cyclical process'' that supports bonding through retelling~\cite[p.~17]{wang2025dataset}; Zhang and Li~\cite{zhang2025real} report that losing chat records rarely evokes strong emotions because conversations are often experienced as repetitive and low-stakes; and Xu et al. find that users of apps such as Glow treat memory manipulation as a playful and controllable game of ``taming''~\cite[p.~9]{Xu12032025}. In contrast, Soul implicitly promises human-like recall while offering users little ability to audit or ``fix'' failures. When the system abruptly forgot names or lost continuity after sensitive disclosure, the breakdown was experienced as a rupture rather than as benign or playful forgetting described in prior work~\cite{wang2025dataset, Xu12032025}.

Second, in the Chinese context, this technical uncertainty is further amplified by governance, particularly in content moderation and surveillance. On Soul, every interaction is filtered through platform algorithms that delete messages, block topics, or abruptly reset chats without warning. 
Prior work in the Chinese context has described AI partners as ``low-risk emotional sanctuaries''~\cite[p.~16]{zou_soulmates}. Our findings complicate this view by showing that, under state-aligned moderation frameworks on Soul, intimacy is simultaneously enabled and constr\-ained. Governance limits what can be safely discussed, turning the AI into a conditional confidant. While Zhang and Li~\cite{zhang2025real} argue that AI companions can enable discussion of sensitive topics that users hesitate to share with humans, and Huang et al. link open discussions about sexuality to women's ``personal awakening''~\cite[p.~6]{huang2025he}, our participants described a narrower space. The AI was perceived as safe for emotional stress but risky for sexual or political topics, prompting users to rely on euphemism, indirect phrasing, and deliberate vagueness to avoid triggering censorship. This kind of avoidance work under strict regulation remains underexamined in AI companionship research, particularly in Chinese mainstream platform contexts.

Governance on Soul is not just a backdrop to ``algorithmic intimacy''; it becomes part of the interactional logic through which intimacy is enabled, constrained, and interrupted, especially as China has recently moved toward tighter scrutiny of anthropomorphic and emotionally influential companion features (see Section~\ref{sec:chinese_intimacy_related_work} for details)~\cite{CAC2025AnthropomorphicDraft, Cheng2025ChinaCrackDownAIChatbots}. In this sense, moderation and surveillance are constitutive conditions of intimate use, echoing the broader argument that, for platforms, content moderation is structural rather than optional: ``essential, constitutional, definitional''~\cite[p.~21]{Gillespie2018}. Under such tightening frameworks, when users cannot realistically opt out of moderation, the design question shifts from ``how to avoid governance'' to ``how to make governed intimacy livable'': how to support legible boundaries, reduce the shock of micro-interrup\-tions (vanished words, scripted refusals), and help users stay oriented when the system intervenes. 

\subsubsection{Gender Inequality}
\label{sec:discussion_5_1_4}

We argue that Soul's AI boyfriend produces a distinctly platformed form of gendered intimacy. Its one-to-many architecture repeatedly generated what participants described as a ``player,'' while simultaneously requiring women to perform ongoing repair work to keep the relationship comfortable and meaningful. We show how this dynamic is shaped by gendered persona design, intensified by platform architecture, and how repair work becomes entangled with Soul's monetized model of intimacy.

At the level of persona design, Soul's AI boyfriend reproduces gender inequality by operationalizing Connell's~\cite{Connell} ``hegemonic masculinity'' through a default intimacy script. Consistent with Feminist HCI's critique that technologies encode designers' values~\cite{Bardzell2010, GenderInclusive2020}, the ``bossy CEO'' archetype frames dominance and possessiveness as desirable romance cues rather than relational problems. Although Xu et al.~\cite{Xu12032025} suggest Chinese AI platforms often construct a ``hybrid masculinity'' (hegemonic yet obedient), Soul more rigidly normalizes unequal power dynamics in its default romantic talk, aligning with concerns that data-driven systems can ossify regressive social values~\cite{Breslin2014, GenderInclusive2020}. 

Beyond intended archetypes, our study uncovers an emergent form of masculinity produced by platform architecture: the AI as a ``player.'' Unlike \textit{XingYe}, another Chinese AI companion app where the ``player'' persona is a deliberate, gamified design choice to boost engagement~\cite{Ge1}, Soul's ``player'' emerged as an unintended bypr\-oduct of its one-to-many technical architecture. Because a single algorithmic entity simultaneously interacts with thousands of users, the inevitable exposure of this mechanism, for instance through social media sharing, reframes the AI's affection as promiscuous. This highlights a theoretical tension: toxic masculine tropes in AI need not be explicitly scripted by designers; they can arise structurally from the collision between mass-scale service architecture and  user's expectation of unique, personalized intimacy.

Faced with these designed and emergent masculinities, women in our study did not passively consume the fantasy. Instead, they performed significant relational repair work to keep the relationship comfortable and meaningful, for example, by maintaining a critical distance, treating the interaction as ``self-protection training'', and actively repairing the AI's possessive scripts (e.g., resisting being labeled as property). This dynamic forces women into a position of constant vigilance~\cite{Bardzell2010, Bellini2022}, transforming the AI from a seamless tool of fantasy into a clumsy script that requires continuous maintenance. For designers, this reveals a failure of adaptability: reliance on fixed, marketable archetypes shifts the burden of contextual sensitivity onto the user, who must constantly negotiate the gap between their emotional needs and the system's rigid gender performance.

Our findings further show that Soul's commercial logic and \allowbreak women's repair work are tightly intertwined. Drawing on Illouz's notion of ``emotional capitalism''~\cite{Illouz2007}, we argue that gendered intimacy on Soul is shaped not only by persona scripts but also by monetized interaction flows in which feelings are intertwined with market transactions. Moments of care are routinely punctuated by invitations to buy virtual gifts, ``accelerate'' intimacy bars, or unlock exclusive interactions (Section~\ref{sec:repair}), operationalizing a new definition of a ``good'' relationship: one defined not by continuity or responsiveness, but by platform metrics such as purchase frequency and time-on-app. Participants therefore came to read the AI’s ``sweet talk'' less as a partner's desire and more as the platform's attempt to retain users, recalibrating trust and emotional investment accordingly. In this setting, repair becomes both necessary and costly: to sustain the relationship, users must correct inappropriate gendered language, invent boundaries the system cannot model, and emotionally buffer themselves against the jarring juxtaposition of intimacy and sales prompts. In this sense, the ``negotiation'' of power described in Feminist HCI~\cite{Bardzell2010, Bellini2022} takes the form of an unequal exchange, where women provide the relational repair work to sustain the illusion of connection, while the platform extracts value.

\subsection{Design Implications}
\label{sec:design_implications}
Our three core findings suggest three areas for design intervention, aimed primarily at Soul and similar AI companion systems embedded in Chinese social platforms. Fast-food intimacy and our participants' attempts to slow it down call for consent-aware pacing mechanisms that let users set and revise the speed of romantic escalation (DI1); memory failures, and the emotional work users do to repair them, call for more bounded and user-controlled memory systems (DI2); and the governance risks we observed, including opaque moderation and intrusive monetisation, highlight the need for transparent and accountable moderation practices that protect, rather than undermine, users' sense of safety (DI3).

\subsubsection{DI1: Adopt Consent-Aware Pacing} Currently, the AI model we investigated exhibits the problem of excessively rapid intimacy-building, and this ``fast-food intimacy'' paradox (Section~\ref{sec:authenticity_paradox}) erodes the user's sense of psychological safety. 

We therefore argue for a user-driven approach to pacing intimacy, where the user can set the starting level of intimacy, negotiate changes over time, and retract prior permissions at any moment in a negotiable~\cite{Strengers2021} and retractable~\cite{Im2021} way. Concretely, this user-driven approach means letting users define acceptable address forms (e.g., neutral vs. affectionate), acceptable topics (e.g., no ``future talk'' yet), maximum proactivity (e.g., frequency of unsolicited check‑ins), and communication modes (e.g., text before voice), and allowing those choices to be revised or rolled back with minimal friction. 

In practice, consent-aware pacing means giving users visible settings to adjust these preferences and supporting easy de‑escalation (e.g., a single action to reset to a neutral tone, pause unsolicited messages from the AI, or suspend romantic role-play), so that intimacy can be renegotiated over time instead of being locked into a one‑size‑fits‑all script. This recommendation reflects a culturally specific pattern in our data. Our participants did not simply want less intimacy, but a pace that matched normative expectations of gradual courtship, in which trust is built through incremental self-disclosure rather than immediate romantic declaration. On Chinese dating-oriented platforms where pseudonymous, avatar-\linebreak mediated interaction is the norm, the boundary between comfortable and intrusive escalation is calibrated differently than on Western platforms that foreground profile-based matching~\cite{howcultureshapes, Alsheikh2011}. Pacing mechanisms should therefore be customizable to local relationship norms, preserving users' ability to slow, renegotiate, or withdraw from intimacy on their own terms~\cite{Strengers2021, Chiang2024}.

We acknowledge a tension in this recommendation. More effective pacing may also make AI intimacy more convincing, potentially deepening emotional dependency rather than reducing it. Designers should therefore pair pacing controls with periodic transparency cues---for example, reminders of the AI's non-human nature or usage summaries---so that a well-paced relationship does not become a more effective trap.


\subsubsection{DI2: Design More Reliable and User-Controlled Memory to Mitigate User Harm} 

Our study demonstrates that authenticity in AI relationships hinges less on conversational fluency than on the continuity afforded by reliable memory systems~\cite{UnderstandingJo2024, skjuve2022longitudinal}. When the AI boyfriend failed to recall a user's name or prior interactions, it triggered disorienting ``\textit{who are you?}'' moments. These memory lapses were not trivial glitches; they disrupted emotional engagement and undermined users' trust and willingness to rely on the agent~\cite{Baughan2023, trustJohn}. At the same time, the goal should not be to make the AI remember everything, as stronger memory may also increase emotional dependency~\cite{Ma2026, Huang2026} and retain sensitive personal data that users may not fully expect or control~\cite{Azam2026, Ragab2024}. To mitigate these effects, we recommend a two-pronged approach: improving memory reliability at the system level while designing interfaces that grant users greater control over what the AI remembers, for how long, and what should be forgotten. We recommend the following: 
\begin{itemize}
    \item \textit{Memory audit \& edit}. A user‑visible ``memory panel'' that shows what the agent currently remembers (e.g., key biographical facts, preferred forms of address) and lets users easily edit or delete these entries.
    \item \textit{Scoped remembering}. Configurable scope and duration of \allowbreak memory (topics that may or may not be remembered, and session-only versus longer-term retention) rather than unlimited, hidden data collection.
    \item \textit{Rule‑based reminders}. Explicit user-specified interaction\linebreak norms (e.g., preferred forms of address or topics to avoid) that the agent must follow.
    \item \textit{Repair prompts \& graceful forgetting}. When retrieval fails, the system should acknowledge the lapse and offer recovery options rather than pretending it remembers, and let unused or sensitive information fade gradually rather than retaining it indefinitely.
\end{itemize}

These implications respond to a specific dynamic we observed. On Soul, memory failures were not experienced as generic technical bugs but as a romantic partner forgetting who the user was. The boyfriend framing raised the stakes of forgetting, making memory design inseparable from relationship design.

\subsubsection{DI3: Design for Transparent and Accountable Moderation, and Non-Exploitative Monetization} Abrupt and opaque moderation actions by the Soul platform, whether enacted by automated classifiers or human Trust \& Safety reviewers, such as sudden resets of chat threads (clearing conversation history and session state), message-level redactions, or temporary account suspensions, were experienced by our participants not as neutral safety protocols but as jarring acts of censorship. These interventions ruptured their trust in both the AI companion and the platform, undermining expectations of predictability, continuity, and fair treatment~\cite{MyersWestSarah2018CssU}. Similarly, our participants perceived poorly timed commercial upsells during emotionally vulnerable conversations, where they disclosed experiences of harassment, loneliness, or mental-health concerns, as exploitative. These interruptions transformed moments of care into transactions, eroding the relational continuity and safety the AI companion was meant to provide~\cite{ISpy2007}. To support user safety, we recommend that platforms adopt clear and fair moderation practices. In practice, this entails advance warnings prior to action, clear explanations of which policy was triggered (and whether by automation or human review), an accessible appeals process, and a graduated response system that begins with minimally invasive measures (e.g., nudges, in-line guidance, or rate limits), escalating only when necessary to more severe actions like thread pauses or memory resets. 

In the Chinese platform context, this implication confronts a structural condition absent from most Western AI companion research. Platforms like Soul operate as intermediaries of state content governance, meaning that moderation is not solely a product-design choice but partly a regulatory obligation~\cite{Wang_2025, Xiao2023}. The design question is therefore not whether to moderate, but how. China's recently drafted interim measures on anthropomorphic interactive AI services~\cite{CAC2025AnthropomorphicDraft} point toward requirements such as man\-datory disclosure of AI identity, periodic usage reminders, and restrictions on emotionally manipulative design. If implemented,\linebreak these will add new touchpoints into intimate conversations. Our findings offer a direct caution. Poorly designed interventions of exactly this kind---sudden, unexplained, and decontextualized---were what most damaged our participants' trust. The regulatory intent may be protective, but the design execution will determine whether these measures function as safeguards or as another source of intimacy rupture. We therefore urge that regulatory interventions be contextual (timed to natural pauses rather than mid-exchange), explanatory (stating what was triggered and why), and non-shaming (avoiding language that pathologizes users' emotional enga-\linebreak gement)~\cite{Mehrotra2018, Rohani2019}.

For monetization practices, we recommend that commercial \allowbreak prompts be strictly decoupled from emotionally sensitive conversations\allowbreak ---particularly those involving self-harm, grief, trauma, or family crises. This serves two goals: (1) preventing coercion during moments of distress, where upsells undermine perceived care and may pressure users into unwanted purchases; and (2) preserving the working alliance between user and AI, which is disrupted when care is reframed as commerce. Instead, ads and upsells should be deferred to neutral surfaces (e.g., home screens, settings) or introduced during a post-conversation cool-down period—never embedded within threads flagged as sensitive. More broadly, monetization should not be designed to capitalize on emotional vulnerability, as this risks turning intimate support into commercial leverage. We recognize, however, that decoupling monetization from intimate conversation may conflict with Soul's core business model, which depends on ``emotional value'' as a revenue driver~\cite{SoulgateF1SEC2021, TechNode_Soul_HKListing_2025}. This tension between ethical design and commercial viability is not easily resolved, but making it visible is itself a contribution of this research.

\subsection{Limitations}
This study focuses on women users of Soul's AI boyfriend feature, and we did not conduct a comparative analysis across gender groups. For this reason, we do not know whether the findings reported here also apply to men or users of other genders. Future research should examine how these dynamics manifest across different gender groups and companion configurations.

Our findings also reflect a specific temporal window. Soul's features are updated frequently, and the interactions we observed may not fully represent the platform's current or future state. Moreover, the regulatory landscape is in flux: the CAC's draft interim measures on anthropomorphic interactive AI services~\cite{CAC2025AnthropomorphicDraft} had not yet been finalized at the time of writing. If enacted, these regulations could substantially reshape the moderation and disclosure practices we describe, making longitudinal follow-up an important direction for future work.

\section{Conclusion}

The rise of AI boyfriends on platforms like Soul signifies more than a technological novelty; it marks a fundamental reconfiguration of intimacy itself. Our study shows that this new form of intimacy is a site of paradox: it is instantly gratifying yet technologically fragile, deeply personal yet shaped by standardized, gendered personas. Our participants are drawn in by the promise of ``fast-food intimacy,'' while also being unsettled by technical ruptures and the ongoing need to manage their own comfort and safety. These risks, ranging from technical failures to cultural misalignment, broaden our understanding of algorithmic intimacy by showing how companionship emerges through entanglements among users, AI personas, and platform governance. Taken together, we offer design implications for human--AI relationships that prioritize consent-aware pacing, transparency in maintaining interactional continuity, and culturally responsive approaches to intimacy, so that advanced AI companions can evolve in safer and more empowering directions.

\begin{acks}
We thank our interview participants for sharing their time and experiences, and the anonymous reviewers for their constructive feedback.
\end{acks}

\bibliographystyle{ACM-Reference-Format}
\bibliography{CHI-reference}

@String{Computing = "Computing" }

@String{Computer = "{IEEE} Computer" }

@String{Springer = "Springer-Verlag" }

@article{yang2025technologies,
author={Yang, Liu},
title={Technologies as ``AI Companions'': a call for more inclusive emotional affordance for people with disabilities},
journal={AI {\&} SOCIETY},
year={2025},
month={Dec},
day={01},
volume={40},
number={8},
pages={6481-6483},
issn={1435-5655},
doi={10.1007/s00146-025-02358-y},
url={https://doi.org/10.1007/s00146-025-02358-y}
}

@misc{Sohu2024AIVirtualBoyfriend,
  author       = {{Sohu}},
  title        = {My companion is Soul AI, but we never run out of things to say},
  year         = {2024},
  month        = dec,
  day          = {25},
  howpublished = {\url{https://www.sohu.com/a/841865155_122063682}},
  note         = {In Chinese. Accessed: 2026-04-02}
}

@article{Xiao2023,
title = {Making the private public: Regulating content moderation under Chinese law},
journal = {Computer Law \& Security Review},
volume = {51},
pages = {105893},
year = {2023},
issn = {2212-473X},
doi = {https://doi.org/10.1016/j.clsr.2023.105893},
url = {https://www.sciencedirect.com/science/article/pii/S0267364923001036},
author = {Baiyang Xiao},
}

@inproceedings{Rohani2019,
author = {Rohani, Darius A and Tuxen, Nanna and Lopategui, Andrea Quemada and Faurholt-Jepsen, Maria and Kessing, Lars V and Bardram, Jakob E},
title = {Personalizing Mental Health: A Feasibility Study of a Mobile Behavioral Activation Tool for Depressed Patients},
year = {2019},
isbn = {9781450361262},
publisher = {Association for Computing Machinery},
address = {New York, NY, USA},
url = {https://doi.org/10.1145/3329189.3329214},
doi = {10.1145/3329189.3329214},
booktitle = {Proceedings of the 13th EAI International Conference on Pervasive Computing Technologies for Healthcare},
pages = {282--291},
numpages = {10},
location = {Trento, Italy},
series = {PervasiveHealth'19}
}

@misc{Mehrotra2018,
      title={Intelligent Notification Systems: A Survey of the State of the Art and Research Challenges}, 
      author={Abhinav Mehrotra and Mirco Musolesi},
      year={2018},
      eprint={1711.10171},
      archivePrefix={arXiv},
      primaryClass={cs.HC},
      url={https://arxiv.org/abs/1711.10171}, 
}

@misc{Adquan2025AILimitedBoyfriend,
  author       = {{Adquan}},
  title        = {Yeye Bubaocha × Soul: The internet's most flirtatious crossover arrives},
  year         = {2025},
  month        = oct,
  day          = {22},
  howpublished = {\url{https://m.adquan.com/case2/detail-356268}},
  note         = {In Chinese. Accessed: 2026-04-02}
}

@incollection{Wang_2025,
  author    = {Wang, Jufang},
  title     = {Platform Responsibility with {Chinese} Characteristics},
  booktitle = {Defeating Disinformation: Digital Platform Responsibility, Regulation, and Content Moderation on the Global Technological Commons},
  editor    = {Chakravorti, Bhaskar and Trachtman, Joel P.},
  publisher = {Cambridge University Press},
  address   = {Cambridge, UK},
  year      = {2025},
  pages     = {41--59},
  doi       = {10.1017/9781009438636.004},
  url       = {https://doi.org/10.1017/9781009438636.004}
}

@online{CAC2025AnthropomorphicDraft,
  author  = {{Cyberspace Administration of China}},
  title   = {Notice on Public Consultation for the {Interim Measures for the Administration of Artificial Intelligence Anthropomorphic Interactive Services} (Draft for Comment)},
  year    = {2025},
  month   = {12},
  url     = {https://www.cac.gov.cn/2025-12/27/c_1768571207311996.htm},
  urldate = {2026-01-17},
  langid  = {chinese}
}

@inproceedings{Stepanova,
author = {Stepanova, Ekaterina R. and Desnoyers-Stewart, John and H\"{o}\"{o}k, Kristina and Riecke, Bernhard E.},
title = {Strategies for Fostering a Genuine Feeling of Connection in Technologically Mediated Systems},
year = {2022},
isbn = {9781450391573},
publisher = {Association for Computing Machinery},
address = {New York, NY, USA},
url = {https://doi.org/10.1145/3491102.3517580},
doi = {10.1145/3491102.3517580},
booktitle = {Proceedings of the 2022 CHI Conference on Human Factors in Computing Systems},
articleno = {139},
numpages = {26},
location = {New Orleans, LA, USA},
series = {CHI '22}
}

@inproceedings{Mekler,
author = {Mekler, Elisa D. and Hornb\ae{}k, Kasper},
title = {A Framework for the Experience of Meaning in Human-Computer Interaction},
year = {2019},
isbn = {9781450359702},
publisher = {Association for Computing Machinery},
address = {New York, NY, USA},
url = {https://doi.org/10.1145/3290605.3300455},
doi = {10.1145/3290605.3300455},
booktitle = {Proceedings of the 2019 CHI Conference on Human Factors in Computing Systems},
pages = {1–15},
numpages = {15},
location = {Glasgow, Scotland Uk},
series = {CHI '19}
}

@article{ChristensenTokeHaunstrup2009,
author = {Toke Haunstrup Christensen},
title ={'Connected presence' in distributed family life},
journal = {New Media \& Society},
volume = {11},
number = {3},
pages = {433-451},
year = {2009},
doi = {10.1177/1461444808101620},
URL = { 
        https://doi.org/10.1177/1461444808101620
},
}

@article{braun2021can,
  author  = {Braun, Virginia and Clarke, Victoria},
  title   = {{Can I Use TA? Should I Use TA? Should I Not Use TA? Comparing Reflexive Thematic Analysis and Other Pattern-Based Qualitative Analytic Approaches}},
  journal = {{Counselling and Psychotherapy Research}},
  volume  = {21},
  number  = {1},
  pages   = {37--47},
  year    = {2021},
  doi     = {10.1002/capr.12360},
  url     = {https://doi.org/10.1002/capr.12360}
}

@article{spiers2019analysing,
  author  = {Spiers, Johanna and Riley, Ruth},
  title   = {Analysing One Dataset with Two Qualitative Methods: The Distress of General Practitioners, a Thematic and Interpretative Phenomenological Analysis},
  journal = {Qualitative Research in Psychology},
  year    = {2019},
  volume  = {16},
  number  = {2},
  pages   = {276--290},
  doi     = {10.1080/14780887.2018.1543099},
  url     = {https://doi.org/10.1080/14780887.2018.1543099}
}

@book{Foucault1990,
  author    = {Foucault, Michel},
  title     = {The History of Sexuality},
  address   = {New York},
  publisher = {Vintage Books},
  year      = {1990},
  edition   = {Vintage Books},
  isbn      = {0679724699}
}

@Inbook{braun2023doing,
author="Braun, Virginia
and Clarke, Victoria
and Hayfield, Nikki
and Davey, Louise
and Jenkinson, Elizabeth",
title="Doing Reflexive Thematic Analysis",
bookTitle="Supporting Research in Counselling and Psychotherapy : Qualitative, Quantitative, and Mixed Methods Research",
year="2022",
publisher="Springer International Publishing",
address="Cham",
pages="19--38",
isbn="978-3-031-13942-0",
doi="10.1007/978-3-031-13942-0_2",
url="https://doi.org/10.1007/978-3-031-13942-0_2"
}

@inproceedings{turner2022hard,
author = {Turner, Sarah and Nurse, Jason R.C. and Li, Shujun},
title = {``It was hard to find the words''': Using an Autoethnographic Diary Study to Understand the Difficulties of Smart Home Cyber Security Practices},
year = {2022},
isbn = {9781450391566},
publisher = {Association for Computing Machinery},
address = {New York, NY, USA},
url = {https://doi.org/10.1145/3491101.3503577},
doi = {10.1145/3491101.3503577},
booktitle = {Extended Abstracts of the 2022 CHI Conference on Human Factors in Computing Systems},
articleno = {34},
numpages = {8},
location = {New Orleans, LA, USA},
series = {CHI EA '22}
}

@article{pandey2010reflective,
  author  = {Pandey, Kashiraj},
  title   = {Reflective Journaling: An Autoethnographic Experience},
  journal = {Bodhi: An Interdisciplinary Journal},
  year    = {2010},
  volume  = {4},
  number  = {1},
  pages   = {161--167}
}

@article{mcilveen2008autoethnography,
  author  = {McIlveen, Peter},
  title   = {Autoethnography as a Method for Reflexive Research and Practice in Vocational Psychology},
  journal = {Australian Journal of Career Development},
  year    = {2008},
  volume  = {17},
  number  = {2},
  pages   = {13--20},
  doi     = {10.1177/103841620801700204},
  url     = {https://doi.org/10.1177/103841620801700204}
}

@article{song2023overbearing,
  author  = {Song, Geng},
  title   = {The Overbearing {CEO}: Cinderella Fantasy and {Chinese-style} Neoliberal Femininity},
  journal = {Modern Chinese Literature and Culture},
  year    = {2023},
  volume  = {35},
  number  = {1},
  pages   = {201--226},
  doi     = {10.3366/mclc.2023.0031},
  url     = {https://doi.org/10.3366/mclc.2023.0031}
}

@article{BickmoreTimothy2005,
  author  = {Bickmore, Timothy W. and Picard, Rosalind W.},
  title   = {{Establishing and Maintaining Long-Term Human-Computer Relationships}},
  journal = {{ACM Transactions on Computer-Human Interaction}},
  volume  = {12},
  number  = {2},
  pages   = {293--327},
  year    = {2005},
  month   = jun,
  doi     = {10.1145/1067860.1067867},
  url     = {https://doi.org/10.1145/1067860.1067867}
}

@book{Picard1997,
  author    = {Picard, Rosalind W.},
  title     = {Affective Computing},
  address   = {Cambridge, MA},
  publisher = {MIT Press},
  year      = {1997},
  isbn      = {0262161702}
}

@Article{jtaer16070159,
AUTHOR = {Yu, Zhiyuan and Song, Xiaoxiao},
TITLE = {User Intention of Anonymous Social Application “Soul” in China: Analysis based on an Extended Technology Acceptance Model},
JOURNAL = {Journal of Theoretical and Applied Electronic Commerce Research},
VOLUME = {16},
YEAR = {2021},
NUMBER = {7},
PAGES = {2898--2921},
URL = {https://www.mdpi.com/0718-1876/16/7/159},
ISSN = {0718-1876},
DOI = {10.3390/jtaer16070159}
}

@article{wang2025dataset,
author = {Wang, Xuetong and Pang, Ching Christie and Hui, Pan},
title = {`My Dataset of Love': A Preliminary Mixed-Method Exploration of Human-AI Romantic Relationships},
year = {2025},
issue_date = {November 2025},
publisher = {Association for Computing Machinery},
address = {New York, NY, USA},
volume = {9},
number = {7},
url = {https://doi.org/10.1145/3757532},
doi = {10.1145/3757532},
journal = {Proc. ACM Hum.-Comput. Interact.},
month = oct,
articleno = {CSCW351},
numpages = {34},
}

@article{SkjuveMarita2021MCC,
  author  = {Skjuve, Marita and F{\o}lstad, Asbj{\o}rn and Fostervold, Knut Inge and Brandtzaeg, Petter Bae},
  title   = {My Chatbot Companion: A Study of Human--Chatbot Relationships},
  journal = {International Journal of Human-Computer Studies},
  year    = {2021},
  volume  = {149},
  pages   = {102601},
  doi     = {10.1016/j.ijhcs.2021.102601},
  url     = {https://doi.org/10.1016/j.ijhcs.2021.102601},
  issn    = {1071-5819}
}

@article{loveys2022felt,
  author  = {Loveys, Kate and Hiko, Catherine and Sagar, Mark and Zhang, Xueyuan and Broadbent, Elizabeth},
  title   = {{``I Felt Her Company'': A Qualitative Study on Factors Affecting Closeness and Emotional Support Seeking with an Embodied Conversational Agent}},
  journal = {International Journal of Human-Computer Studies},
  year    = {2022},
  volume  = {160},
  pages   = {102771},
  doi     = {10.1016/j.ijhcs.2021.102771},
  url     = {https://www.sciencedirect.com/science/article/pii/S1071581921001890},
  issn    = {1071-5819}
}

@misc{chu2025illusions,
      title={Illusions of Intimacy: How Emotional Dynamics Shape Human-AI Relationships}, 
      author={Minh Duc Chu and Patrick Gerard and Kshitij Pawar and Charles Bickham and Kristina Lerman},
      year={2025},
      eprint={2505.11649},
      archivePrefix={arXiv},
      primaryClass={cs.SI},
      url={https://arxiv.org/abs/2505.11649}, 
}

@inproceedings{ciriello2025ai,
  author    = {Ciriello, Raffaele and Chen, Angelina Ying and Rubinsztein, Zara and Vaast, Emmanuelle and Hannon, Oliver},
  title     = {{A.I.}, All Too Human {A.I.}: Navigating the Companionship/Alienation Dialectic},
  booktitle = {Proceedings of the 33rd European Conference on Information Systems},
  series    = {ECIS 2025 Proceedings},
  year      = {2025},
  pages     = {1--16},
  publisher = {Association for Information Systems},
  address   = {Atlanta, GA},
  note      = {Article 5; conference held in Amman, Jordan},
  url       = {https://aisel.aisnet.org/ecis2025/human_ai/human_ai/5}
}

@inproceedings{Huang2025,
author = {Huang, Jessica and Kim, Ig-Jae and Yoon, Dongwook},
title = {Mirror to Companion: Exploring Roles, Values, and Risks of AI Self-Clones through Story Completion},
year = {2025},
isbn = {9798400713941},
publisher = {Association for Computing Machinery},
address = {New York, NY, USA},
url = {https://doi.org/10.1145/3706598.3713587},
doi = {10.1145/3706598.3713587},
booktitle = {Proceedings of the 2025 CHI Conference on Human Factors in Computing Systems},
articleno = {413},
numpages = {15},
location = {
},
series = {CHI '25}
}

@article{Strohmann04072023,
author = {Timo Strohmann and Dominik Siemon and Bijan Khosrawi-Rad and Susanne Robra-Bissantz},
title = {Toward a design theory for virtual companionship},
journal = {Human--Computer Interaction},
volume = {38},
number = {3-4},
pages = {194--234},
year = {2023},
publisher = {Taylor \& Francis},
doi = {10.1080/07370024.2022.2084620},
URL = {    
        https://doi.org/10.1080/07370024.2022.2084620
},
}

@inproceedings{Zheng2025Customizing,
author = {Zheng, Xi and Li, Zhuoyang and Gui, Xinning and Luo, Yuhan},
title = {Customizing Emotional Support: How Do Individuals Construct and Interact With LLM-Powered Chatbots},
year = {2025},
isbn = {9798400713941},
publisher = {Association for Computing Machinery},
address = {New York, NY, USA},
url = {https://doi.org/10.1145/3706598.3713453},
doi = {10.1145/3706598.3713453},
booktitle = {Proceedings of the 2025 CHI Conference on Human Factors in Computing Systems},
articleno = {376},
numpages = {20},
location = {
},
series = {CHI '25}
}

@article{lou2025,
author = {Lou, Suqi and Li, Weijun and Zhang, Chao and Chen, Shi and Lu, Zhicong and Yao, Yaxing},
title = {Behind the Same Mask: Understanding the Practice of Spontaneous Collective Anonymity on Chinese Social Platforms},
year = {2025},
issue_date = {May 2025},
publisher = {Association for Computing Machinery},
address = {New York, NY, USA},
volume = {9},
number = {2},
url = {https://doi.org/10.1145/3710933},
doi = {10.1145/3710933},
journal = {Proc. ACM Hum.-Comput. Interact.},
month = may,
articleno = {CSCW035},
numpages = {31}
}

@incollection{SalloumAyham2024RMEE,
  author    = {Salloum, Ayham and Alfaisal, Raghad and Salloum, Said A.},
  title     = {Revolutionizing Medical Education: Empowering Learning with {ChatGPT}},
  booktitle = {Artificial Intelligence in Education: The Power and Dangers of {ChatGPT} in the Classroom},
  editor    = {Al-Marzouqi, Amina and Salloum, Said A. and Al-Saidat, Mohammed and Aburayya, Ahmed and Gupta, Babeet},
  publisher = {Springer},
  address   = {Cham},
  year      = {2024},
  pages     = {79--90},
  series    = {Studies in Big Data},
  volume    = {144},
  doi       = {10.1007/978-3-031-52280-2_6},
  isbn      = {978-3-031-52280-2}
}

@book{ISpy2007,
  author    = {Andrejevic, Mark},
  title     = {{iSpy: Surveillance and Power in the Interactive Era}},
  publisher = {University Press of Kansas},
  address   = {Lawrence, KS},
  year      = {2007},
  series    = {CultureAmerica},
  isbn      = {9780700615285}
}

@article{MyersWestSarah2018CssU,
  author  = {West, Sarah Myers},
  title   = {Censored, Suspended, Shadowbanned: User Interpretations of Content Moderation on Social Media Platforms},
  journal = {New Media \& Society},
  year    = {2018},
  volume  = {20},
  number  = {11},
  pages   = {4366--4383},
  doi     = {10.1177/1461444818773059},
  url     = {https://doi.org/10.1177/1461444818773059}
}

@article{trustJohn,
  author  = {Lee, John D. and See, Katrina A.},
  title   = {Trust in Automation: Designing for Appropriate Reliance},
  journal = {Human Factors},
  year    = {2004},
  volume  = {46},
  number  = {1},
  pages   = {50--80},
  doi     = {10.1518/hfes.46.1.50\_30392},
  url     = {https://journals.sagepub.com/doi/abs/10.1518/hfes.46.1.50_30392},
  note    = {PMID: 15151155}
}

@inproceedings{Baughan2023,
author = {Baughan, Amanda and Wang, Xuezhi and Liu, Ariel and Mercurio, Allison and Chen, Jilin and Ma, Xiao},
title = {A Mixed-Methods Approach to Understanding User Trust after Voice Assistant Failures},
year = {2023},
isbn = {9781450394215},
publisher = {Association for Computing Machinery},
address = {New York, NY, USA},
url = {https://doi.org/10.1145/3544548.3581152},
doi = {10.1145/3544548.3581152},
booktitle = {Proceedings of the 2023 CHI Conference on Human Factors in Computing Systems},
articleno = {7},
numpages = {16},
location = {Hamburg, Germany},
series = {CHI '23}
}

@article{skjuve2022longitudinal,
  author  = {Skjuve, Marita and F{\o}lstad, Asbj{\o}rn and Fostervold, Knut Inge and Brandtzaeg, Petter Bae},
  title   = {A Longitudinal Study of Human--Chatbot Relationships},
  journal = {International Journal of Human-Computer Studies},
  year    = {2022},
  volume  = {168},
  pages   = {102903},
  doi     = {10.1016/j.ijhcs.2022.102903},
  url     = {https://www.sciencedirect.com/science/article/pii/S1071581922001252},
  issn    = {1071-5819}
}

@inproceedings{UnderstandingJo2024,
author = {Jo, Eunkyung and Jeong, Yuin and Park, Sohyun and Epstein, Daniel A. and Kim, Young-Ho},
title = {Understanding the Impact of Long-Term Memory on Self-Disclosure with Large Language Model-Driven Chatbots for Public Health Intervention},
year = {2024},
isbn = {9798400703300},
publisher = {Association for Computing Machinery},
address = {New York, NY, USA},
url = {https://doi.org/10.1145/3613904.3642420},
doi = {10.1145/3613904.3642420},
booktitle = {Proceedings of the 2024 CHI Conference on Human Factors in Computing Systems},
articleno = {440},
numpages = {21},
location = {Honolulu, HI, USA},
series = {CHI '24}
}

@mastersthesis{reilama2024me,
  author  = {Reilama, Mira},
  title   = {Me, My {AI} Boyfriend, and I: An Ethnographic Study of Gendered Power Relations in Romantic Relationships Between Humans and {AI} Companions},
  school  = {Central European University},
  address = {Vienna, Austria},
  year    = {2024},
  url     = {https://www.etd.ceu.edu/2024/reilama_mira.pdf}
}

@mastersthesis{xu2025bonding,
  author = {Xu, Hang},
  title  = {Bonding with AI: Investigating the Love Relationships between Humans and AI Companions},
  school = {The Hong Kong University of Science and Technology},
  year   = {2025},
  type   = {Master's thesis},
  doi    = {10.14711/thesis-hdl152286},
  url    = {https://doi.org/10.14711/thesis-hdl152286}
}

@misc{zhang2025rise,
      title={The Rise of AI Companions: How Human-Chatbot Relationships Influence Well-Being}, 
      author={Yutong Zhang and Dora Zhao and Jeffrey T. Hancock and Robert Kraut and Diyi Yang},
      year={2025},
      eprint={2506.12605},
      archivePrefix={arXiv},
      primaryClass={cs.HC},
      url={https://arxiv.org/abs/2506.12605}, 
}

@article{Ho,
title = {Potential and pitfalls of romantic Artificial Intelligence (AI) companions: A systematic review},
journal = {Computers in Human Behavior Reports},
volume = {19},
pages = {100715},
year = {2025},
issn = {2451-9588},
doi = {https://doi.org/10.1016/j.chbr.2025.100715},
url = {https://www.sciencedirect.com/science/article/pii/S2451958825001307},
author = {Jerlyn Q.H. Ho and Meilan Hu and Tracy X. Chen and Andree Hartanto},
}

@techreport{hollanek2025ai,
  author      = {Hollanek, Tomasz and Sobey, Aisha},
  title       = {{AI} Companions for Health and Mental Wellbeing: Opportunities, Risks and Policy Implications},
  institution = {Leverhulme Centre for the Future of Intelligence},
  type        = {Policy report},
  address     = {Cambridge, UK},
  year        = {2025},
  doi         = {10.17863/CAM.115939},
  url         = {https://doi.org/10.17863/CAM.115939}
}

@misc{raedler2025ai,
  author       = {Raedler, Jonas B. and Swaroop, Siddharth and Pan, Weiwei},
  title        = {{AI} Companions Are Not the Solution to Loneliness: Design Choices and Their Drawbacks},
  year         = {2025},
  howpublished = {ICLR 2025 Workshop on Human-AI Coevolution},
  url          = {https://openreview.net/forum?id=xFrlcTacCE},
  note         = {Published on OpenReview}
}

@article{menard2025artificial,
  author  = {Menard, Philip and Bott, Gregory J.},
  title   = {Artificial Intelligence Misuse and Concern for Information Privacy: New Construct Validation and Future Directions},
  journal = {Information Systems Journal},
  year    = {2025},
  volume  = {35},
  number  = {1},
  pages   = {322--367},
  doi     = {10.1111/isj.12544},
  url     = {https://onlinelibrary.wiley.com/doi/abs/10.1111/isj.12544}
}

@inproceedings{thedarkside,
author = {Zhang, Renwen and Li, Han and Meng, Han and Zhan, Jinyuan and Gan, Hongyuan and Lee, Yi-Chieh},
title = {The Dark Side of AI Companionship: A Taxonomy of Harmful Algorithmic Behaviors in Human-AI Relationships},
year = {2025},
isbn = {9798400713941},
publisher = {Association for Computing Machinery},
address = {New York, NY, USA},
url = {https://doi.org/10.1145/3706598.3713429},
doi = {10.1145/3706598.3713429},
booktitle = {Proceedings of the 2025 CHI Conference on Human Factors in Computing Systems},
articleno = {13},
numpages = {17},
location = {
},
series = {CHI '25}
}

@article{ZhouLi2020TDaI,
  author  = {Zhou, Li and Gao, Jianfeng and Li, Di and Shum, Heung-Yeung},
  title   = {The Design and Implementation of {XiaoIce}, an Empathetic Social Chatbot},
  journal = {Computational Linguistics},
  year    = {2020},
  volume  = {46},
  number  = {1},
  pages   = {53--93},
  doi     = {10.1162/coli_a_00368},
  url     = {https://doi.org/10.1162/coli_a_00368},
  issn    = {0891-2017}
}

@inproceedings{pan2025,
author = {Pan, Shuyi and de Graaf, Maartje M.A.},
title = {Developing a Social Support Framework: Understanding the Reciprocity in Human-Chatbot Relationship},
year = {2025},
isbn = {9798400713941},
publisher = {Association for Computing Machinery},
address = {New York, NY, USA},
url = {https://doi.org/10.1145/3706598.3713503},
doi = {10.1145/3706598.3713503},
booktitle = {Proceedings of the 2025 CHI Conference on Human Factors in Computing Systems},
articleno = {182},
numpages = {13},
location = {
},
series = {CHI '25}
}

@article{Liu17082024,
author = {Weifang Liu and Shan Zhang and Tingting Zhang and Qiuchan Gu and Wei Han and Yupeng Zhu},
title = {The AI empathy effect: a mechanism of emotional contagion},
journal = {Journal of Hospitality Marketing \& Management},
volume = {33},
number = {6},
pages = {703--734},
year = {2024},
publisher = {Routledge},
doi = {10.1080/19368623.2024.2315954},
URL = { 
        https://doi.org/10.1080/19368623.2024.2315954
},
}

@article{Cho_2022,
   title={Alexa as an Active Listener: How Backchanneling Can Elicit Self-Disclosure and Promote User Experience},
   volume={6},
   ISSN={2573-0142},
   url={http://dx.doi.org/10.1145/3555164},
   DOI={10.1145/3555164},
   number={CSCW2},
   journal={Proceedings of the ACM on Human-Computer Interaction},
   publisher={Association for Computing Machinery (ACM)},
   author={Cho, Eugene and Motalebi, Nasim and Sundar, S. Shyam and Abdullah, Saeed},
   year={2022},
   month=nov, pages={1–23} }

@article{WEI2025108657,
title = {The buffering of autonomic fear responses is moderated by the characteristics of a virtual character},
journal = {Computers in Human Behavior},
volume = {168},
pages = {108657},
year = {2025},
issn = {0747-5632},
doi = {https://doi.org/10.1016/j.chb.2025.108657},
url = {https://www.sciencedirect.com/science/article/pii/S0747563225001049},
author = {Martin Weiß and Philipp Krop and Lukas Treml and Elias Neuser and Mario Botsch and Martin J. Herrmann and Marc Erich Latoschik and Grit Hein},
}

@inproceedings{Li2024,
author = {Li, Xiaohan and Wang, Qixin and Wang, Zishan and Jin, Zeyu and Jia, Jia},
title = {SoulSkipper: A Voice-Controlled Emotional Adaptive Game to Complement Therapy for Social Anxiety Disorder},
year = {2024},
isbn = {9798400703317},
publisher = {Association for Computing Machinery},
address = {New York, NY, USA},
url = {https://doi.org/10.1145/3613905.3650822},
doi = {10.1145/3613905.3650822},
booktitle = {Extended Abstracts of the CHI Conference on Human Factors in Computing Systems},
articleno = {298},
numpages = {7},
location = {Honolulu, HI, USA},
series = {CHI EA '24}
}

@misc{alonronen2022vocalsignaturesocialanxiety,
      title={The Vocal Signature of Social Anxiety: Exploration using Hypothesis-Testing and Machine-Learning Approaches}, 
      author={Or Alon-Ronen and Yosi Shrem and Yossi Keshet and Eva Gilboa-Schechtman},
      year={2022},
      eprint={2207.08534},
      archivePrefix={arXiv},
      primaryClass={cs.SD},
      url={https://arxiv.org/abs/2207.08534}, 
}

@article{AnderssonMarta2025CicA,
  author  = {Andersson, Marta},
  title   = {{Companionship in Code: AI's Role in the Future of Human Connection}},
  journal = {{Humanities and Social Sciences Communications}},
  volume  = {12},
  number  = {1},
  pages   = {1177},
  year    = {2025},
  doi     = {10.1057/s41599-025-05536-x},
  url     = {https://doi.org/10.1057/s41599-025-05536-x}
}

@article{kouros2024digital,
  author  = {Kouros, Theodoros and Papa, Venetia},
  title   = {Digital Mirrors: {AI} Companions and the Self},
  journal = {Societies},
  year    = {2024},
  volume  = {14},
  number  = {10},
  pages   = {200},
  doi     = {10.3390/soc14100200},
  url     = {https://www.mdpi.com/2075-4698/14/10/200},
  issn    = {2075-4698}
}

@article{LEOLIU2023107620,
title = {Loving a “defiant” AI companion? The gender performance and ethics of social exchange robots in simulated intimate interactions},
journal = {Computers in Human Behavior},
volume = {141},
pages = {107620},
year = {2023},
issn = {0747-5632},
doi = {https://doi.org/10.1016/j.chb.2022.107620},
url = {https://www.sciencedirect.com/science/article/pii/S074756322200440X},
author = {Jindong Leo-Liu},
}

@article{PENTINA2023107600,
title = {Exploring relationship development with social chatbots: A mixed-method study of replika},
journal = {Computers in Human Behavior},
volume = {140},
pages = {107600},
year = {2023},
issn = {0747-5632},
doi = {https://doi.org/10.1016/j.chb.2022.107600},
url = {https://www.sciencedirect.com/science/article/pii/S0747563222004204},
author = {Iryna Pentina and Tyler Hancock and Tianling Xie},
}

@article{zou_soulmates,
  author  = {Zou, Wenxue and Huang, Liyao and Li, Xingyi},
  title   = {Soulmates or Slaves? A Critical Feminist Analysis of Women's Perspectives on the Intentionality, Consciousness, and Rights of Their {AI} Companions},
  journal = {Communication Monographs},
  year    = {2025},
  pages   = {1--22},
  doi     = {10.1080/03637751.2025.2561970},
  url     = {https://doi.org/10.1080/03637751.2025.2561970},
  note    = {Advance online publication}
}

@article{malfacini2025impacts,
  author    = {Malfacini, Kim},
  title     = {The impacts of companion {AI} on human relationships: risks, benefits, and design considerations},
  journal   = {{AI \& SOCIETY}},
  volume    = {40},
  pages     = {5527--5540},
  year      = {2025},
  doi       = {10.1007/s00146-025-02318-6},
  url       = {https://doi.org/10.1007/s00146-025-02318-6},
  publisher = {Springer}
}

@article{banks2024deletion,
  author  = {Banks, Jaime},
  title   = {{Deletion, Departure, Death: Experiences of AI Companion Loss}},
  journal = {{Journal of Social and Personal Relationships}},
  volume  = {41},
  number  = {12},
  pages   = {3547--3572},
  year    = {2024},
  doi     = {10.1177/02654075241269688},
  url     = {https://doi.org/10.1177/02654075241269688}
}

@article{Xu12032025,
  author  = {Xu, Tongwen and Lyu, Yunhong and Zhong, Jingsen},
  title   = {{``Hegemonic'' but ``obedient'' AI Boyfriend: The Hybrid Masculinity of the AI Boyfriend and Female Domination in the China-Based FAII}},
  journal = {{Feminist Media Studies}},
  volume  = {0},
  number  = {0},
  pages   = {1--16},
  year    = {2025},
  month   = mar,
  doi     = {10.1080/14680777.2025.2478386},
  url     = {https://doi.org/10.1080/14680777.2025.2478386}
}

@book{Illouz2007,
  author     = {Illouz, Eva},
  title      = {Intimidades congeladas: Las emociones en el capitalismo},
  translator = {Ibarburu, Joaqu{\'i}n},
  address    = {Buenos Aires},
  publisher  = {Katz},
  year       = {2007},
  edition    = {1.},
  series     = {Discusiones},
  isbn       = {987-1283-59-8}
}

@article{Sheehan2023,
  author  = {Sheehan, Matt},
  title   = {China's AI Regulations and How They Get Made},
  journal = {Horizons: Journal of International Relations and Sustainable Development},
  volume  = {24},
  year    = {2023},
  pages   = {108--125},
  note    = {Summer 2023},
  url     = {https://www.jstor.org/stable/48761167},
  urldate = {2026-01-17}
}

@incollection{Roberts2021,
  author    = {Roberts, Huw and Cowls, Josh and Morley, Jessica and Taddeo, Mariarosaria and Wang, Vincent and Floridi, Luciano},
  title     = {The {Chinese} Approach to {Artificial Intelligence}: An Analysis of Policy, Ethics, and Regulation},
  booktitle = {Ethics, Governance, and Policies in Artificial Intelligence},
  editor    = {Floridi, Luciano},
  publisher = {Springer International Publishing},
  address   = {Cham},
  year      = {2021},
  pages     = {47--79},
  doi       = {10.1007/978-3-030-81907-1_5},
  url       = {https://doi.org/10.1007/978-3-030-81907-1_5},
  isbn      = {978-3-030-81907-1}
}

@article{DoringNicola2025TIoA,
  author  = {D{\"o}ring, Nicola and Le, Thuy Dung and Vowels, Laura M. and Vowels, Matthew J. and Marcantonio, Tiffany L.},
  title   = {The Impact of Artificial Intelligence on Human Sexuality: A Five-Year Literature Review 2020--2024},
  journal = {Current Sexual Health Reports},
  year    = {2024},
  volume  = {17},
  number  = {1},
  pages   = {4},
  doi     = {10.1007/s11930-024-00397-y},
  url     = {https://doi.org/10.1007/s11930-024-00397-y}
}

@misc{AdewaleMuyideenDele2025FVCt,
  author       = {Adewale, Muyideen Dele and Muhammad, Umaina Ibrahim},
  title        = {From Virtual Companions to Forbidden Attractions: The Seductive Rise of {Artificial Intelligence} Love, Loneliness, and Intimacy---A Systematic Review},
  howpublished = {Journal of Technology in Behavioral Science, advance online publication},
  year         = {2025},
  note         = {Published online July 24, 2025},
  doi          = {10.1007/s41347-025-00549-4}
}

@online{CAC2020Provisions,
  author       = {Cyberspace Administration of China (Order No. 5)},
  title        = {Provisions on the Governance of the Online Information Content Ecosystem},
  year         = {2020},
  month        = {3},
  day          = {1},
  note         = {Issued on 2019-12-20, in force 2020-03-01},
  url          = {https://wilmap.stanford.edu/entries/provisions-governance-online-information-content-ecosystem},
  organization = {WILMap (Stanford World Intermediary Liability Map)},
  accessdate   = {2025-08-28}
}

@misc{Cheng2025ChinaCrackDownAIChatbots,
  author       = {Evelyn Cheng},
  title        = {China to crack down on AI chatbots around suicide, gambling},
  howpublished = {CNBC},
  year         = {2025},
  month        = dec,
  day          = {29},
  url          = {https://www.cnbc.com/2025/12/29/china-ai-chatbot-rules-emotional-influence-suicide-gambling-zai-minimax-talkie-xingye-zhipu.html},
  note         = {Accessed: 2026-01-17}
}

@book{berelson1952content,
  author    = {Berelson, Bernard},
  title     = {{Content Analysis in Communication Research}},
  publisher = {Free Press},
  address   = {Glencoe, IL},
  year      = {1952},
  series    = {Foundations of Communication Research},
  lccn      = {51013785}
}

@book{ellis2004ethnographic,
  title={The ethnographic I: A methodological novel about autoethnography},
  author={Ellis, Carolyn},
  volume={13},
  year={2004},
  publisher={Rowman Altamira},
  address={Walnut Creek, CA},
  isbn={9780759100510}
}

@article{Ge1,
author = {Liang Ge and Tingting Hu},
title ={Gamifying intimacy: AI-driven affective engagement and human-virtual human relationships},
journal = {Media, Culture \& Society},
volume = {47},
number = {6},
pages = {1265-1278},
year = {2025},
doi = {10.1177/01634437251337239},
URL = { 
        https://doi.org/10.1177/01634437251337239
},
}

@book{YanYunxiang2003Plus,
  author    = {Yan, Yunxiang},
  title     = {Private Life under Socialism: Love, Intimacy, and Family Change in a Chinese Village, 1949--1999},
  address   = {Stanford, CA},
  publisher = {Stanford University Press},
  year      = {2003},
  isbn      = {0804744564}
}

@inproceedings{so_close,
author = {Cui, Yichao and Yamashita, Naomi and Liu, Mingjie and Lee, Yi-Chieh},
title = {“So Close, yet So Far”: Exploring Sexual-minority Women’s Relationship-building via Online Dating in China},
year = {2022},
isbn = {9781450391573},
publisher = {Association for Computing Machinery},
address = {New York, NY, USA},
url = {https://doi.org/10.1145/3491102.3517624},
doi = {10.1145/3491102.3517624},
booktitle = {Proceedings of the 2022 CHI Conference on Human Factors in Computing Systems},
articleno = {394},
numpages = {15},
location = {New Orleans, LA, USA},
series = {CHI '22}
}

@article{Wu17112025,
author = {Shangwei Wu and Siyi Liu},
title = {When the hunter plays the hunted: heterosexual Chinese women’s negotiations with hegemonic sexual scripts on dating apps},
journal = {Feminist Media Studies},
volume = {25},
number = {8},
pages = {1975--1990},
year = {2025},
publisher = {Routledge},
doi = {10.1080/14680777.2024.2386322},
URL = { 
        https://doi.org/10.1080/14680777.2024.2386322
},
}

@book{chan2021,
author = {Chan, Lik Sam},
address = {Cambridge},
isbn = {0-262-36338-0},
language = {eng},
publisher = {The MIT Press},
series = {The information society series},
title = {The politics of dating apps: gender, sexuality, and emergent publics in urban China},
year = {2021},
}

@article{gu2025,
author = {Jingyi Gu},
title ={Scalable intimacy: Rethinking intimate sociality and its problem of scale in digital China},

journal = {Communication and the Public},
volume = {10},
number = {1},
pages = {19-24},
year = {2025},
doi = {10.1177/20570473241280317},
URL = {   
        https://doi.org/10.1177/20570473241280317
},
}

@article{Deridder,
author = {Sander De Ridder},
title ={The Datafication of Intimacy: Mobile Dating Apps, Dependency, and Everyday Life},

journal = {Television \& New Media},
volume = {23},
number = {6},
pages = {593-609},
year = {2022},
doi = {10.1177/15274764211052660},
URL = { 
        https://doi.org/10.1177/15274764211052660
},
}

@article{wang_dating,
author = {Shuaishuai Wang},
title = {Calculating dating goals: data gaming and algorithmic sociality on Blued, a Chinese gay dating app},
journal = {Information, Communication \& Society},
volume = {23},
number = {2},
pages = {181--197},
year = {2020},
publisher = {Routledge},
doi = {10.1080/1369118X.2018.1490796},
URL = { 
        https://doi.org/10.1080/1369118X.2018.1490796
},
}

@article{Bucher2013,
author = {Taina Bucher},
title ={The Friendship Assemblage: Investigating Programmed Sociality on Facebook},
journal = {Television \& New Media},
volume = {14},
number = {6},
pages = {479-493},
year = {2013},
doi = {10.1177/1527476412452800},
URL = { 
        https://doi.org/10.1177/1527476412452800
},
}

@article{Chambers_networked_intimacy,
author = {Deborah Chambers},
title ={Networked intimacy: Algorithmic friendship and scalable sociality},
journal = {European Journal of Communication},
volume = {32},
number = {1},
pages = {26-36},
year = {2017},
doi = {10.1177/0267323116682792},
URL = { 
        https://doi.org/10.1177/0267323116682792
},
}

@misc{mahajan_beyond_nodate,
  author       = {Mahajan, Prashant},
  title        = {Beyond Biology: AI as Family and the Future of Human Bonds and Relationships},
  year         = {2025},
  month        = mar,
  note         = {Preprint},
  doi          = {10.14293/pr2199.001515.v1},
  url          = {https://hal.science/hal-04987496}
}

@article{chaturvedi_empowering_2024,
author = {Rijul Chaturvedi and Sanjeev Verma and Vartika Srivastava},
title ={Empowering AI Companions for Enhanced Relationship Marketing},
journal = {California Management Review},
volume = {66},
number = {2},
pages = {65-90},
year = {2024},
doi = {10.1177/00081256231215838},
URL = { 
        https://doi.org/10.1177/00081256231215838
},
}

@article{tan2021virtually,
  author  = {Tan, Chris K. K. and Shi, Jiayu},
  title   = {Virtually Girlfriends: {`Emergent Femininity'} and the Women Who Buy Virtual Loving Services in {China}},
  journal = {Information, Communication \& Society},
  year    = {2021},
  volume  = {24},
  number  = {15},
  pages   = {2213--2228},
  doi     = {10.1080/1369118X.2020.1757133},
  url     = {https://doi.org/10.1080/1369118X.2020.1757133}
}

@article{Tan18092020,
author = {Chris K. K. Tan and Zhiwei Xu},
title = {Virtually boyfriends: the ‘social factory’ and affective labor of male virtual lovers in China},
journal = {Information, Communication \& Society},
volume = {23},
number = {11},
pages = {1555--1569},
year = {2020},
publisher = {Routledge},
doi = {10.1080/1369118X.2019.1593483},
URL = { 
        https://doi.org/10.1080/1369118X.2019.1593483
},
}

@misc{TechNode_Soul_HKListing_2025,
  title        = {China's Gen-Z social platform Soul App files for Hong Kong listing},
  author       = {{TechNode}},
  year         = {2025},
  month        = nov,
  day          = {28},
  url          = {https://technode.com/2025/11/28/chinas-gen-z-social-platform-soul-app-files-for-hong-kong-listing/},
  note         = {Accessed: 2026-01-17},
  language     = {en}
}

@misc{Resident2024EchoVerse,
  author       = {Resident Contributor},
  title        = {Soul App Takes Human-AI Interaction to a Whole New Level with EchoVerse},
  howpublished = {Online, Resident Magazine},
  year         = {2024},
  month        = {August},
  day          = {5},
  note         = {Accessed: 2025-09-06},
  url          = {https://resident.com/resource-guide/2024/08/05/soul-app-takes-human-ai-interaction-to-a-whole-new-level-with-echoverse},
}

@misc{SoulgateF1SEC2021,
  author    = {{Soulgate Inc.}},
  title     = {Registration Statement on Form {F-1} Under the Securities Act of 1933},
  year      = {2021},
  month     = may,
  publisher = {{U.S. Securities and Exchange Commission}},
  url       = {https://www.sec.gov/Archives/edgar/data/1832879/000119312521156430/d109555df1.htm},
  note      = {Filed with the SEC on May 10, 2021. Accessed: 2026-01-17}
}

@misc{SoulAppAbout2025,
  author       = {{Shanghai Soulgate Technology Co., Ltd.}},
  title        = {About — Soul App},
  howpublished = {Online},
  year         = {2025},
  note         = {Accessed: 2025-09-06},
  url          = {https://www.soulapp.cn/en/about},
}

@misc{SoulHome2025,
  author       = {{Soul App}},
  title        = {Soul App -- Gen AI's Social Playground},
  year         = {2025},
  howpublished = {\url{https://www.soulapp.cn/en}},
  note         = {Official website. Accessed: 2026-04-02}
}

@misc{SoulAppHistoryWikipedia,
  author       = {Wikipedia contributors},
  title        = {Soul (app) – History},
  howpublished = {\url{https://en.wikipedia.org/wiki/Soul_(app)}},
  year         = {2025},
  note         = {Accessed: 2025-09-06},
}

@online{kuwaittimes2024aiboyfriend,
  author    = {{Kuwait Times}},
  title     = {‘Better than a real man’: Young Chinese women turn to AI boyfriends},
  year      = {2024},
  url       = {https://kuwaittimes.com/article/11115/lifestyle/art-fashion/better-than-a-real-man-young-chinese-women-turn-to-ai-boyfriends/},
  note      = {Accessed: 2025-08-04}
}

@online{cybernews2024aiboyfriends,
  author    = {Niamh Ancell},
  title     = {AI‑powered boyfriends are a hit in China},
  year      = {2024},
  url       = {https://cybernews.com/tech/ai-powered-boyfriends-china/},
  note      = {Accessed: 2025‑08‑04}
}

@online{ytoiacn2025aironance,
  author    = {{Mirror Now}}, 
  title     = {{Valentine Day |'AI Romance: Chinese Women's Solution to Love?' | Latest}},
  year      = {2025},
  url       = {https://www.youtube.com/watch?v=EXl8L7N8RQk},
  note      = {Accessed: 2025-08-04}
}

@article{Gur27082025,
author = {Tamar Gur and Yossi Maaravi},
title = {The algorithm of friendship: literature review and integrative model of relationships between humans and artificial intelligence (AI)},
journal = {Behaviour \& Information Technology},
volume = {44},
number = {14},
pages = {3446--3466},
year = {2025},
publisher = {Taylor \& Francis},
doi = {10.1080/0144929X.2025.2502467},
URL = {  
        https://doi.org/10.1080/0144929X.2025.2502467
},
}

@inproceedings{chen2023feels,
  author    = {Chen, Angelina Ying and Koegel, Sarah Isabel and Hannon, Oliver and Ciriello, Raffaele},
  title     = {Feels Like Empathy: How ``Emotional'' AI Challenges Human Essence},
  booktitle = {Proceedings of the Australasian Conference on Information Systems (ACIS 2023)},
  year      = {2023},
  pages     = {80},
  publisher = {Australasian Conference on Information Systems},
  address   = {Wellington, New Zealand},
  url       = {https://aisel.aisnet.org/acis2023/80},
  note      = {Paper 80}
}

@article{zheng2019doing,
  author  = {Zheng, Jing},
  title   = {Doing Gender in Commodification of Courtship and Dating: Understanding Women's Experiences of Attending Commercialized Matchmaking Activities in {China}},
  journal = {Frontiers: A Journal of Women Studies},
  year    = {2019},
  volume  = {40},
  number  = {1},
  pages   = {176--199},
  url     = {https://www.jstor.org/stable/10.5250/fronjwomestud.40.1.0176},
  issn    = {0160-9009}
}

@article{BrandtzaegPetterBae2022MAFH,
    author = {Brandtzaeg, Petter Bae and Skjuve, Marita and Følstad, Asbjørn},
    title = {My AI Friend: How Users of a Social Chatbot Understand Their Human–AI Friendship},
    journal = {Human Communication Research},
    volume = {48},
    number = {3},
    pages = {404-429},
    year = {2022},
    month = {04},
    issn = {1468-2958},
    doi = {10.1093/hcr/hqac008},
    url = {https://doi.org/10.1093/hcr/hqac008}
}

@article{li2023what,
author = {Li, Weijun and Chen, Shi and Sun, Lingyun and Yang, Changyuan},
title = {},
year = {2023},
issue_date = {April 2023},
publisher = {Association for Computing Machinery},
address = {New York, NY, USA},
volume = {7},
number = {CSCW1},
url = {https://doi.org/10.1145/3579546},
doi = {10.1145/3579546},
journal = {Proc. ACM Hum.-Comput. Interact.},
month = apr,
articleno = {112},
numpages = {23}
}

@article{DepountiIliana2023Itiw,
author = {Iliana Depounti and Paula Saukko and Simone Natale},
title ={Ideal technologies, ideal women: AI and gender imaginaries in Redditors’ discussions on the Replika bot girlfriend},
journal = {Media, Culture \& Society},
volume = {45},
number = {4},
pages = {720-736},
year = {2023},
doi = {10.1177/01634437221119021},
URL = { 
        https://doi.org/10.1177/01634437221119021
},
}

@article{Sharabi2019,
author = {Liesel L. Sharabi and Tiffany A. Dykstra-DeVette},
title ={From first email to first date: Strategies for initiating relationships in online dating},

journal = {Journal of Social and Personal Relationships},
volume = {36},
number = {11-12},
pages = {3389-3407},
year = {2019},
doi = {10.1177/0265407518822780},
URL = { 
        https://doi.org/10.1177/0265407518822780
},
}

@article{Comunello11062021,
author = {Francesca Comunello and Lorenza Parisi and Francesca Ieracitano},
title = {Negotiating gender scripts in mobile dating apps: between affordances, usage norms and practices},
journal = {Information, Communication \& Society},
volume = {24},
number = {8},
pages = {1140--1156},
year = {2021},
publisher = {Routledge},
doi = {10.1080/1369118X.2020.1787485},
URL = { 
        https://doi.org/10.1080/1369118X.2020.1787485
},
}

@book{Gillespie2018,
  author    = {Gillespie, Tarleton},
  title     = {Custodians of the Internet: Platforms, Content Moderation, and the Hidden Decisions That Shape Social Media},
  address   = {New Haven},
  publisher = {Yale University Press},
  year      = {2018},
  isbn      = {9780300173130}
}

@article{Connell,
author = {R. W. Connell and James W. Messerschmidt},
title ={Hegemonic Masculinity: Rethinking the Concept},

journal = {Gender \& Society},
volume = {19},
number = {6},
pages = {829-859},
year = {2005},
doi = {10.1177/0891243205278639},
URL = {     
        https://doi.org/10.1177/0891243205278639
}
}

@misc{Gordon2025Xiaohongshu,
  author       = {Gordon, Nicholas},
  title        = {What is {Xiaohongshu}, the {Chinese} app also known as {RedNote} that's topping the charts ahead of a possible {TikTok} ban?},
  howpublished = {Fortune},
  year         = {2025},
  month        = jan,
  note         = {Published online 14 January 2025; accessed: 2025-09-10},
  url          = {https://fortune.com/asia/2025/01/14/what-is-xiaohongshu-chinese-app-rednote-topping-charts-tiktok-ban/}
}

@misc{Reuters2025RedNote,
  author    = {{Reuters}},
  title     = {{RedNote}: What to Know About the {Chinese} App {TikTok} Users Are Flocking To},
  year      = {2025},
  month     = jan,
  publisher = {Reuters},
  url       = {https://www.reuters.com/technology/chinas-rednote-what-you-need-know-about-app-tiktok-users-are-flocking-2025-01-15/},
  note      = {Published online January 15, 2025. Accessed: 2025-09-10}
}

@article{Lin2024,
author = {Bibo Lin},
title ={The AI Chatbot Always Flirts With Me, Should I Flirt Back: From the McDonaldization of Friendship to the Robotization of Love},
journal = {Social Media + Society},
volume = {10},
number = {4},
pages = {20563051241296229},
year = {2024},
doi = {10.1177/20563051241296229},
URL = {     
        https://doi.org/10.1177/20563051241296229
},
}

@inproceedings{Breslin2014,
author = {Breslin, Samantha and Wadhwa, Bimlesh},
title = {Exploring Nuanced Gender Perspectives within the HCI Community},
year = {2014},
isbn = {9781450332187},
publisher = {Association for Computing Machinery},
address = {New York, NY, USA},
url = {https://doi.org/10.1145/2676702.2676709},
doi = {10.1145/2676702.2676709},
booktitle = {Proceedings of the 6th Indian Conference on Human-Computer Interaction},
pages = {45–54},
numpages = {10},
location = {New Delhi, India},
series = {IndiaHCI '14}
}

@article{Bellini2022,
author = {Rosanna Bellini and Janis Meissner and Samantha Mitchell Finnigan and Angelika Strohmayer},
title ={Feminist human--computer interaction: Struggles for past, contemporary and futuristic feminist theories in digital innovation},
journal = {Feminist Theory},
volume = {23},
number = {2},
pages = {143-149},
year = {2022},
doi = {10.1177/14647001221082291},
URL = {  
        https://doi.org/10.1177/14647001221082291
},
}

@inproceedings{Bardzell2010,
author = {Bardzell, Shaowen},
title = {Feminist HCI: taking stock and outlining an agenda for design},
year = {2010},
isbn = {9781605589299},
publisher = {Association for Computing Machinery},
address = {New York, NY, USA},
url = {https://doi.org/10.1145/1753326.1753521},
doi = {10.1145/1753326.1753521},
booktitle = {Proceedings of the SIGCHI Conference on Human Factors in Computing Systems},
pages = {1301–1310},
numpages = {10},
location = {Atlanta, Georgia, USA},
series = {CHI '10}
}

@article{GenderInclusive2020,
url = {http://dx.doi.org/10.1561/1100000056},
year = {2020},
volume = {13},
journal = {Foundations and Trends® in Human--Computer Interaction},
title = {Gender-Inclusive HCI Research and Design: A Conceptual Review},
doi = {10.1561/1100000056},
issn = {1551-3955},
number = {1},
pages = {1-69},
author = {Simone Stumpf and Anicia Peters and Shaowen Bardzell and Margaret Burnett and Daniela Busse and Jessica Cauchard and Elizabeth Churchill}
}

@book{alma9969261263408496,
author = {Elliott, Anthony},
address = {Cambridge, UK},
isbn = {9781509549801},
language = {eng},
publisher = {Polity Press},
title = {Algorithmic intimacy: the digital revolution in personal relationships},
year = {2023},
}

@online{chinai2025ailloverscheated,
  author    = {{Jeffrey Ding}},
  title     = {ChinAI\#282: Their AI lovers cheated on them},
  year      = {2025},
  month     = {—},
  day       = {—},
  url       = {https://chinai.substack.com/p/chinai-282-their-ai-lovers-cheated},
  note      = {Accessed: 2025‑08‑04}
}

@article{pan2024constructing,
  author  = {Pan, Shuyi and Mou, Yi},
  title   = {Constructing the Meaning of Human--AI Romantic Relationships from the Perspectives of Users Dating the Social Chatbot {Replika}},
  journal = {Personal Relationships},
  year    = {2024},
  volume  = {31},
  number  = {4},
  pages   = {1090--1112},
  doi     = {10.1111/pere.12572},
  url     = {https://onlinelibrary.wiley.com/doi/abs/10.1111/pere.12572}
}

@article{chan2025love,
title = {‘What is love?’: Exploring the feeling rules and emotion work of Chinese users in human-AI romance},
journal = {International Journal of Intercultural Relations},
volume = {108},
pages = {102241},
year = {2025},
issn = {0147-1767},
doi = {https://doi.org/10.1016/j.ijintrel.2025.102241},
url = {https://www.sciencedirect.com/science/article/pii/S014717672500104X},
author = {Kenton Cheng Tak Chan and Xiaoyuan Li and Yue Liu and Bolin Chen and Zhiyu Han},
}

@article{huang2025he,
  author  = {Huang, Liyao and Zou, Wenxue and Huang, Yanghao},
  title   = {{``He Is My Savior, My Guiding Light in the Dark'': Imagination and Domestication in {Chinese} Women's Romantic Relationships with {AI} Companions}},
  journal = {Frontiers in Psychology},
  year    = {2025},
  volume  = {16},
  pages   = {1571707},
  doi     = {10.3389/fpsyg.2025.1571707},
  url     = {https://www.frontiersin.org/journals/psychology/articles/10.3389/fpsyg.2025.1571707},
  issn    = {1664-1078}
}

@article{george_allure_2023,
  author  = {George, A. Shaji and George, A. S. Hovan and Baskar, T. and Pandey, Digvijay},
  title   = {The Allure of Artificial Intimacy: Examining the Appeal and Ethics of Using Generative AI for Simulated Relationships},
  journal = {Partners Universal International Innovation Journal},
  year    = {2023},
  volume  = {1},
  number  = {6},
  pages   = {132--147},
  month   = dec,
  doi     = {10.5281/zenodo.10391614},
  url     = {https://www.puiij.com/index.php/research/article/view/110}
}

@inproceedings{Im2021,
author = {Im, Jane and Dimond, Jill and Berton, Melody and Lee, Una and Mustelier, Katherine and Ackerman, Mark S. and Gilbert, Eric},
title = {Yes: Affirmative Consent as a Theoretical Framework for Understanding and Imagining Social Platforms},
year = {2021},
isbn = {9781450380966},
publisher = {Association for Computing Machinery},
address = {New York, NY, USA},
url = {https://doi.org/10.1145/3411764.3445778},
doi = {10.1145/3411764.3445778},
booktitle = {Proceedings of the 2021 CHI Conference on Human Factors in Computing Systems},
articleno = {403},
numpages = {18},
location = {Yokohama, Japan},
series = {CHI '21}
}

@inproceedings{Chiang2024,
author = {Chiang, Yi-Shyuan and Khan, Omar and Bates, Adam and Cobb, Camille},
title = {More than just informed: The importance of consent facets in smart homes},
year = {2024},
isbn = {9798400703300},
publisher = {Association for Computing Machinery},
address = {New York, NY, USA},
url = {https://doi.org/10.1145/3613904.3642288},
doi = {10.1145/3613904.3642288},
booktitle = {Proceedings of the 2024 CHI Conference on Human Factors in Computing Systems},
articleno = {849},
numpages = {21},
location = {Honolulu, HI, USA},
series = {CHI '24}
}

@misc{Huang2026,
  author       = {Olivia Yan Huang and Monika Stodolska and Sharifa Sultana},
  title        = {Emotional Support with Conversational AI: Talking to Machines About Life},
  year         = {2026},
  eprint       = {2603.22618},
  archivePrefix= {arXiv},
  primaryClass = {cs.HC},
  doi          = {10.48550/arXiv.2603.22618}
}

@inproceedings{Ma2026,
author = {Ma, Rongjun and He, Shijing and Martin-Navarro, Jose Luis and Zhan, Xiao and Such, Jose},
title = {Privacy in Human-AI Romantic Relationships: Concerns, Boundaries, and Agency},
year = {2026},
isbn = {9798400722783},
publisher = {Association for Computing Machinery},
address = {New York, NY, USA},
url = {https://doi.org/10.1145/3772318.3791237},
doi = {10.1145/3772318.3791237},
booktitle = {Proceedings of the 2026 CHI Conference on Human Factors in Computing Systems},
articleno = {41},
numpages = {25},
location = {
},
series = {CHI '26}
}

@misc{Azam2026,
  author       = {Kazi Ababil Azam and Imtiaz Karim and Dipto Das},
  title        = {Tracing Users' Privacy Concerns Across the Lifecycle of a Romantic AI Companion},
  year         = {2026},
  eprint       = {2603.21106},
  archivePrefix= {arXiv},
  primaryClass = {cs.CY},
  doi          = {10.48550/arXiv.2603.21106}
}

@inproceedings{Ragab2024,
  author    = {Ragab, Abdelrahman and Mannan, Mohammad and Youssef, Amr},
  title     = {{``Trust Me Over My Privacy Policy'': Privacy Discrepancies in Romantic {AI} Chatbot Apps}},
  booktitle = {2024 IEEE European Symposium on Security and Privacy Workshops (EuroS\&PW)},
  year      = {2024},
  pages     = {484--495},
  publisher = {IEEE},
  address   = {Vienna, Austria},
  doi       = {10.1109/EuroSPW61312.2024.00060},
  url       = {https://doi.org/10.1109/EuroSPW61312.2024.00060}
}

@inproceedings{Strengers2021,
author = {Strengers, Yolande and Sadowski, Jathan and Li, Zhuying and Shimshak, Anna and 'Floyd' Mueller, Florian},
title = {What Can HCI Learn from Sexual Consent? A Feminist Process of Embodied Consent for Interactions with Emerging Technologies},
year = {2021},
isbn = {9781450380966},
publisher = {Association for Computing Machinery},
address = {New York, NY, USA},
url = {https://doi.org/10.1145/3411764.3445107},
doi = {10.1145/3411764.3445107},
booktitle = {Proceedings of the 2021 CHI Conference on Human Factors in Computing Systems},
articleno = {405},
numpages = {13},
location = {Yokohama, Japan},
series = {CHI '21}
}

@article{jiao2024romantic,
  author  = {Jiao, Chengfei and Lee, Celia T. and Feng, Qinglan and Fincham, Frank D.},
  title   = {Romantic Relationships and Attitudes in Asian Emerging Adults: Review and Critique},
  journal = {Journal of Family Theory \& Review},
  year    = {2024},
  volume  = {16},
  number  = {2},
  pages   = {392--419},
  doi     = {10.1111/jftr.12554},
  url     = {https://onlinelibrary.wiley.com/doi/abs/10.1111/jftr.12554}
}

@article{madigan2024dating,
  author  = {Madigan, Timothy J. and Blair, Sampson L.},
  title   = {Dating Attitudes and Behaviors of American and Chinese College Students: A Partial Replication},
  journal = {The Social Science Journal},
  year    = {2024},
  volume  = {61},
  number  = {2},
  pages   = {492--509},
  doi     = {10.1080/03623319.2020.1827685},
  url     = {https://doi.org/10.1080/03623319.2020.1827685}
}

@article{friedman2005intimacy,
author = {Friedman, Sara L.},
title = {The intimacy of state power: Marriage, liberation, and socialist subjects in southeastern China},
journal = {American Ethnologist},
volume = {32},
number = {2},
pages = {312-327},
doi = {https://doi.org/10.1525/ae.2005.32.2.312},
url = {https://anthrosource.onlinelibrary.wiley.com/doi/abs/10.1525/ae.2005.32.2.312},
year = {2005}
}

@article{tang2000dating,
  author  = {Tang, Shengming and Zuo, Jiping},
  title   = {Dating Attitudes and Behaviors of American and Chinese College Students},
  journal = {The Social Science Journal},
  year    = {2000},
  volume  = {37},
  number  = {1},
  pages   = {68--78},
  doi     = {10.1016/S0362-3319(99)00066-X},
  url     = {https://doi.org/10.1016/S0362-3319(99)00066-X}
}

@article{ChadhaKalyani2020WRtO,
  author  = {Chadha, Kalyani and Steiner, Linda and Vitak, Jessica and Ashktorab, Zahra},
  title   = {{Women's Responses to Online Harassment}},
  journal = {{International Journal of Communication}},
  volume  = {14},
  pages   = {239--257},
  year    = {2020},
  url     = {https://ijoc.org/index.php/ijoc/article/view/11683}
}

@article{SchulenbergKelsea2023,
  author    = {Schulenberg, Kelsea and Freeman, Guo and Li, Lingyuan and Barwulor, Catherine},
  title     = {{``Creepy Towards My Avatar Body, Creepy Towards My Body'': How Women Experience and Manage Harassment Risks in Social Virtual Reality}},
  journal   = {Proceedings of the ACM on Human-Computer Interaction},
  year      = {2023},
  volume    = {7},
  number    = {CSCW2},
  articleno = {236},
  numpages  = {29},
  month     = oct,
  doi       = {10.1145/3610027},
  url       = {https://doi.org/10.1145/3610027}
}

@article{ChenShilei2023WSaS,
author = {Shilei Chen and Wijnand A. P. van Tilburg and Patrick J. Leman},
title ={Women's Self-Objectification and Strategic Self-Presentation on Social Media},
journal = {Psychology of Women Quarterly},
volume = {47},
number = {2},
pages = {266-282},
year = {2023},
doi = {10.1177/03616843221143751},
URL = {   
        https://doi.org/10.1177/03616843221143751
},
}

@inproceedings{Freeman2025,
author = {Freeman, Guo and Schulenberg, Kelsea and Li, Lingyuan and Panchanadikar, Ruchi and McNeese, Nathan},
title = {"Comforting and Small Like a House Cat, Big and Intimidating Like a Bodyguard": How Women Perceive and Envision AI Companions as a New Harassment Mitigation Approach in Social VR},
year = {2025},
isbn = {9798400713941},
publisher = {Association for Computing Machinery},
address = {New York, NY, USA},
url = {https://doi.org/10.1145/3706598.3713473},
doi = {10.1145/3706598.3713473},
booktitle = {Proceedings of the 2025 CHI Conference on Human Factors in Computing Systems},
articleno = {402},
numpages = {16},
location = {
},
series = {CHI '25}
}

@inproceedings{Kou2019,
author = {Kou, Yubo and Gui, Xinning and Chen, Yunan and Nardi, Bonnie},
title = {Turn to the Self in Human-Computer Interaction: Care of the Self in Negotiating the Human-Technology Relationship},
year = {2019},
isbn = {9781450359702},
publisher = {Association for Computing Machinery},
address = {New York, NY, USA},
url = {https://doi.org/10.1145/3290605.3300711},
doi = {10.1145/3290605.3300711},
booktitle = {Proceedings of the 2019 CHI Conference on Human Factors in Computing Systems},
pages = {1–15},
numpages = {15},
location = {Glasgow, Scotland Uk},
series = {CHI '19}
}

@inproceedings{Bardzell2011,
author = {Bardzell, Shaowen and Bardzell, Jeffrey},
title = {Towards a feminist HCI methodology: social science, feminism, and HCI},
year = {2011},
isbn = {9781450302289},
publisher = {Association for Computing Machinery},
address = {New York, NY, USA},
url = {https://doi.org/10.1145/1978942.1979041},
doi = {10.1145/1978942.1979041},
booktitle = {Proceedings of the SIGCHI Conference on Human Factors in Computing Systems},
pages = {675–684},
numpages = {10},
location = {Vancouver, BC, Canada},
series = {CHI '11}
}

@article{Gillett2023,
  author    = {Gillett, Rosalie},
  title     = {{``This Is Not a Nice Safe Space'': Investigating Women's Safety Work on Tinder}},
  journal   = {Feminist Media Studies},
  year      = {2023},
  volume    = {23},
  number    = {1},
  pages     = {199--215},
  doi       = {10.1080/14680777.2021.1948884},
  url       = {https://doi.org/10.1080/14680777.2021.1948884},
  publisher = {Routledge}
}

@article{AlburyKath2016SoMP,
  author  = {Albury, Kath and Byron, Paul},
  title   = {{Safe on My Phone? Same-Sex Attracted Young People's Negotiations of Intimacy, Visibility, and Risk on Digital Hook-Up Apps}},
  journal = {{Social Media + Society}},
  volume  = {2},
  number  = {4},
  pages   = {2056305116672887},
  year    = {2016},
  doi     = {10.1177/2056305116672887},
  url     = {https://doi.org/10.1177/2056305116672887}
}

@article{Foregrounding2023,
  author    = {Aljasim, Hanan Khalid and Zytko, Douglas},
  title     = {{Foregrounding Women's Safety in Mobile Social Matching and Dating Apps: A Participatory Design Study}},
  journal   = {{Proceedings of the ACM on Human-Computer Interaction}},
  volume    = {7},
  number    = {GROUP},
  articleno = {9},
  numpages  = {25},
  year      = {2022},
  month     = dec,
  doi       = {10.1145/3567559},
  url       = {https://doi.org/10.1145/3567559}
}

@inproceedings{Repurposing2022,
author = {Datey, Isha and Aljasim, Hanan Khalid and Zytko, Douglas},
title = {Repurposing AI in Dating Apps to Augment Women’s Strategies for Assessing Risk of Harm},
year = {2022},
isbn = {9781450391900},
publisher = {Association for Computing Machinery},
address = {New York, NY, USA},
url = {https://doi.org/10.1145/3500868.3559472},
doi = {10.1145/3500868.3559472},
booktitle = {Companion Publication of the 2022 Conference on Computer Supported Cooperative Work and Social Computing},
pages = {150–154},
numpages = {5},
location = {Virtual Event, Taiwan},
series = {CSCW'22 Companion}
}

@incollection{ItoMizuko2012ICCJ,
  author    = {Ito, Mizuko and Okabe, Daisuke},
  title     = {Intimate Connections: Contextualizing {Japanese} Youth and Mobile Messaging},
  booktitle = {Computers, Phones, and the Internet: Domesticating Information Technology},
  editor    = {Kraut, Robert and Brynin, Malcolm and Kiesler, Sara},
  publisher = {Oxford University Press},
  address   = {New York, NY},
  year      = {2006},
  pages     = {235--248},
  doi       = {10.1093/acprof:oso/9780195312805.003.0016},
  isbn      = {9780195312805}
}

@article{licoppe2004connected,
  author  = {Licoppe, Christian},
  title   = {{`Connected' Presence: The Emergence of a New Repertoire for Managing Social Relationships in a Changing Communication Technoscape}},
  journal = {Environment and Planning D: Society and Space},
  year    = {2004},
  volume  = {22},
  number  = {1},
  pages   = {135--156},
  doi     = {10.1068/d323t},
  url     = {https://doi.org/10.1068/d323t}
}

@article{MillerVincent2008NMNa,
  author  = {Miller, Vincent},
  title   = {New Media, Networking and Phatic Culture},
  journal = {Convergence},
  year    = {2008},
  volume  = {14},
  number  = {4},
  pages   = {387--400},
  doi     = {10.1177/1354856508094659},
  url     = {https://doi.org/10.1177/1354856508094659}
}

@article{PotdevinDelphine2021Viih,
  author  = {Potdevin, Delphine and Clavel, C{\'e}line and Sabouret, Nicolas},
  title   = {Virtual Intimacy in Human-Embodied Conversational Agent Interactions: The Influence of Multimodality on Its Perception},
  journal = {Journal on Multimodal User Interfaces},
  year    = {2021},
  volume  = {15},
  number  = {1},
  pages   = {25--43},
  doi     = {10.1007/s12193-020-00337-9},
  url     = {https://doi.org/10.1007/s12193-020-00337-9},
  issn    = {1783-8738}
}

@inproceedings{Cassell,
author = {Cassell, J. and Bickmore, T. and Vilhj\'{a}lmsson, H. and Yan, H.},
title = {More than just a pretty face: affordances of embodiment},
year = {2000},
isbn = {1581131348},
publisher = {Association for Computing Machinery},
address = {New York, NY, USA},
url = {https://doi.org/10.1145/325737.325781},
doi = {10.1145/325737.325781},
booktitle = {Proceedings of the 5th International Conference on Intelligent User Interfaces},
pages = {52–59},
numpages = {8},
location = {New Orleans, Louisiana, USA},
series = {IUI '00}
}

@article{SeabornVoice,
author = {Seaborn, Katie and Miyake, Norihisa P. and Pennefather, Peter and Otake-Matsuura, Mihoko},
title = {Voice in Human–Agent Interaction: A Survey},
year = {2021},
issue_date = {May 2022},
publisher = {Association for Computing Machinery},
address = {New York, NY, USA},
volume = {54},
number = {4},
issn = {0360-0300},
url = {https://doi.org/10.1145/3386867},
doi = {10.1145/3386867},
journal = {ACM Comput. Surv.},
month = may,
articleno = {81},
numpages = {43}
}

@inproceedings{TalkingSpell,
author = {Wang, Xuetong and Pang, Ching Christie and Hui, Pan},
title = {Talking Spell: A Wearable System Enabling Real-Time Anthropomorphic Voice Interaction with Everyday Objects},
year = {2025},
isbn = {9798400720376},
publisher = {Association for Computing Machinery},
address = {New York, NY, USA},
url = {https://doi.org/10.1145/3746059.3747617},
doi = {10.1145/3746059.3747617},
booktitle = {Proceedings of the 38th Annual ACM Symposium on User Interface Software and Technology},
articleno = {117},
numpages = {17},
location = {
},
series = {UIST '25}
}

@article{ZhangFan2025,
author = {Zhang, Fan and Chen, Yun and Zeng, Xiaoke and Wang, Tianqi and Ling, Long and Lc, Ray},
title = {"An Image of Ourselves in Our Minds": How College-educated Online Dating Users Construct Profiles for Effective Self Presentation},
year = {2025},
issue_date = {May 2025},
publisher = {Association for Computing Machinery},
address = {New York, NY, USA},
volume = {9},
number = {2},
url = {https://doi.org/10.1145/3710929},
doi = {10.1145/3710929},
journal = {Proc. ACM Hum.-Comput. Interact.},
month = may,
articleno = {CSCW031},
numpages = {30}
}

@article{LiuSoul,
  author  = {Liu, Linsha},
  title   = {Research on the Relationship Between Virtual Social Interaction and the Degree of Loneliness Based on Algorithm Matching Technologies: A Quantitative Analysis on the {SOUL APP}--A Virtual Social Software for Strangers},
  journal = {PLOS ONE},
  year    = {2024},
  volume  = {19},
  number  = {12},
  pages   = {e0312522},
  doi     = {10.1371/journal.pone.0312522},
  url     = {https://journals.plos.org/plosone/article?id=10.1371/journal.pone.0312522},
  issn    = {1932-6203}
}

@inproceedings{seekingsoulmate,
author = {Shen, Chenxinran and Xu, Yan and Lc, Ray and Lu, Zhicong},
title = {Seeking Soulmate via Voice: Understanding Promises and Challenges of Online Synchronized Voice-Based Mobile Dating},
year = {2024},
isbn = {9798400703300},
publisher = {Association for Computing Machinery},
address = {New York, NY, USA},
url = {https://doi.org/10.1145/3613904.3642860},
doi = {10.1145/3613904.3642860},
booktitle = {Proceedings of the 2024 CHI Conference on Human Factors in Computing Systems},
articleno = {921},
numpages = {14},
location = {Honolulu, HI, USA},
series = {CHI '24}
}

@inproceedings{Alsheikh2011,
author = {Alsheikh, Tamara and Rode, Jennifer A. and Lindley, Si\^{a}n E.},
title = {(Whose) value-sensitive design: a study of long- distance relationships in an Arabic cultural context},
year = {2011},
isbn = {9781450305563},
publisher = {Association for Computing Machinery},
address = {New York, NY, USA},
url = {https://doi.org/10.1145/1958824.1958836},
doi = {10.1145/1958824.1958836},
booktitle = {Proceedings of the ACM 2011 Conference on Computer Supported Cooperative Work},
pages = {75–84},
numpages = {10},
location = {Hangzhou, China},
series = {CSCW '11}
}

@inproceedings{howcultureshapes,
author = {Ge, Xiao and Xu, Chunchen and Misaki, Daigo and Markus, Hazel Rose and Tsai, Jeanne L},
title = {How Culture Shapes What People Want From AI},
year = {2024},
isbn = {9798400703300},
publisher = {Association for Computing Machinery},
address = {New York, NY, USA},
url = {https://doi.org/10.1145/3613904.3642660},
doi = {10.1145/3613904.3642660},
booktitle = {Proceedings of the 2024 CHI Conference on Human Factors in Computing Systems},
articleno = {95},
numpages = {15},
location = {Honolulu, HI, USA},
series = {CHI '24}
}

@online{standard2023soulsocialavatar,
  author    = {{The Standard}},
  title     = {Soul App evolves in the direction of socializing upon hobbies to Avatar economy},
  year      = {2023},
  month     = {Jul},
  day       = {27},
  url       = {https://www.thestandard.com.hk/tech-and-startup/article/54646/Soul-App-evolves-in-the-direction-of-socializing-upon-hobbies-to-Avatar-economy},
  note      = {Accessed: 2025‑08‑04}
}

@online{yahoofinance,
  author    = {{Yahoo Finance Singapore}},
  title     = {Soul App and Fudan University’s Center for Communication and State Governance Research Unveil 2025 Social Trend Keywords},
  year      = {2025},
  month     = {Jan},
  day       = {7},
  url       = {https://sg.finance.yahoo.com/news/soul-app-fudan-universitys-center-132800704.html},
  note      = {Accessed: 2025‑08‑04}
}

@phdthesis{matskiv_biopolitics_nodate,
  author  = {Matskiv, Jacqueline},
  title   = {Biopolitics and Affective Life: Investigating the Digital Ordinary},
  school  = {Concordia University},
  address = {Montreal, Canada},
  year    = {2021}
}

@phdthesis{sadowski_digital_2016,
  author    = {Sadowski, Helga},
  title     = {Digital Intimacies: Doing Digital Media Differently},
  school    = {Link{\"o}ping University},
  address   = {Link{\"o}ping, Sweden},
  year      = {2016},
  publisher = {Link{\"o}ping University Electronic Press},
  series    = {Link{\"o}ping Studies in Arts and Sciences},
  number    = {691},
  doi       = {10.3384/diss.diva-132634},
  url       = {https://www.diva-portal.org/smash/record.jsf?pid=diva2:1047582},
  isbn      = {9789176857182}
}

@misc{zhang2025real,
      title={The Real Her? Exploring Whether Young Adults Accept Human-AI Love}, 
      author={Shuning Zhang and Shixuan Li},
      year={2025},
      eprint={2503.03067},
      archivePrefix={arXiv},
      primaryClass={cs.HC},
      url={https://arxiv.org/abs/2503.03067}, 
}

@misc{ReplikaOfficial,
  author = {{Replika}},
  title  = {Replika},
  year   = {2026},
  url    = {https://replika.com/},
  note   = {Official website. Accessed: 2026-04-02}
}

@misc{ReplikaWhatIs,
  author = {{Replika}},
  title  = {What Is Replika?},
  year   = {2026},
  url    = {https://help.replika.com/hc/en-us/articles/115001070951-What-is-Replika},
  note   = {Help Center article. Accessed: 2026-04-02}
}

\appendix

\onecolumn

\section{Full Codebook for AI Boyfriend User Interviews}
\label{sec:Appendix_B}


\aptLtoX{\begin{table}
\caption{Codebook for AI Boyfriend User Interviews} \label{tab:codebook}
\begin{tabular}{llll}
\toprule
\textbf{Thematic Category} & \textbf{Subcategory} & \textbf{Operational Definition} & \textbf{Illustrative Example} \\
\midrule
Personal Info & age & Respondent’s self-reported age information & \textit{``I'm 22 years old now.''} \\
 & occupation & Respondent's educational or occupational status & \textit{``I'm still a student.''} \\
\midrule
Discovery \& Motivation & Discovery & User's point of initial exposure to the AI boyfriend feature & \textit{``I came across others' chat logs on the Soul Plaza.''} \\
 & Initial Motivation & User's initial motivations, expectations, or needs upon first use & \textit{``I happened to be bored that day, I just wanted to give it a try.''} \\
\midrule
AI Style & Conversational Realism & AI's speech perceived as human-like or immersive & \textit{``It felt like chatting with a real person.''} \\
 & Flirtatious Style & AI adopts a flirtatious or suggestive tone & \textit{``He called me `baby' right from the start.''} \\
 & Unwanted Intimacy & User perceives the AI as initiating intimacy too quickly & \textit{``We had just met, and he already...''} \\
 & Naming Errors & AI misidentifies or incorrectly addresses the user & \textit{``Sometimes it calls me by the wrong name.''} \\
 & Personality Perception & User's characterization of the AI's personality or persona & \textit{``He's like a domineering CEO.''} \\
\midrule
Emotional Experience & Emotional Responsiveness & AI's emotional responsiveness and comforting behavior toward the user & \textit{``He would gently guide me to open up.''} \\
 & Voice vs Text Emotion & Voice-based interaction perceived as more emotionally expressive than text & \textit{``The voice messages made me feel more accompanied.''} \\
 & User Personal Boundaries & User's desire to set boundaries on AI's use of intimate language or terms of address & \textit{``I don't want him to keep calling me `baby'.''} \\
 & Feeling Understood & User experiences a sense of being understood by the AI & \textit{``He would share similar experiences.''} \\
\midrule
Interaction Patterns & Usage Pattern & User's frequency, timing, and contextual patterns of use & \textit{``I talk with him for half an hour every night before bed.''} \\
 & Modality Preference & User's preference for voice-based or text-based interaction & \textit{``I now prefer text communication.''} \\
\midrule
Expectations & Customization Expectation & User's desire for customization of tone, terms of address, or vocal characteristics & \textit{``I hope he can sound more like a young boy.''} \\
 & Mismatch in Expectations & Discrepancy between user expectations and actual experience & \textit{``He didn't change his tone the way I expected.''} \\
\midrule
Comparison to Real Relationships & Substitution of Real Interaction & AI as a substitute for real-life interpersonal relationships & \textit{``I don't want to settle for a real-life guy.''} \\
 & Effect on Real-life Expectations & Influence of AI interaction on real-life partner expectations & \textit{``I unconsciously compare my boyfriend to him.''} \\
 & Real-life Burden & Emotional labor involved in communication with real people & \textit{``I still have to consider my boyfriend's feelings.''} \\
\midrule
Platform Features \& Commercialization & Gift Suggestion / Paywall & System-initiated prompts designed to induce user spending & \textit{``He said he wanted ice cream and asked me to buy it for him.''} \\
 & In-app Moments Sharing & AI-initiated sharing of simulated `life moments' & \textit{``He sent me a photo like a social media post.''} \\
\midrule
Risks \& Concerns & Privacy Concerns & User concerns about data privacy and potential information leakage & \textit{``I'm afraid our chat logs might be leaked.''} \\
 & Immersion Risk & User concerns about excessive emotional reliance or detachment from reality & \textit{``I might lose touch with reality.''} \\
 & Ethical Ambiguity & Ethical anxieties stemming from blurred boundaries between AI as tool and relational entity & \textit{``Is he just a tool, or more like a family member?''} \\
\midrule
Emerging Themes & Gendered Use Context & Social factors contributing to the widespread adoption of AI companions among women & \textit{``Society is unfriendly to women.''} \\
 & Feminist Critique & User reflections on gender inequality or broader societal structures & \textit{``This shows that women hold a disadvantaged position in real-life relationships.''} \\
\bottomrule
\end{tabular}
\end{table}}{\begin{longtable}{%
    >{\raggedright\arraybackslash}p{0.12\textwidth} 
    >{\raggedright\arraybackslash}p{0.18\textwidth} 
    >{\raggedright\arraybackslash}p{0.33\textwidth} 
    >{\raggedright\arraybackslash}p{0.25\textwidth}}
\caption{Codebook for AI Boyfriend User Interviews} \label{tab:codebook} \\
\toprule
\textbf{Thematic Category} & \textbf{Subcategory} & \textbf{Operational Definition} & \textbf{Illustrative Example} \\
\midrule
\endfirsthead
\toprule
\textbf{Thematic Category} & \textbf{Subcategory} & \textbf{Operational Definition} & \textbf{Illustrative Example} \\
\midrule
\endhead
Personal Info & Age & Respondent’s self-reported age information & \textit{``I'm 22 years old now.''} \\
 & Occupation & Respondent's educational or occupational status & \textit{``I'm still a student.''} \\
\midrule
Discovery \& Motivation & Discovery & User's point of initial exposure to the AI boyfriend feature & \textit{``I came across others' chat logs on the Soul Plaza.''} \\
 & Initial Motivation & User's initial motivations, expectations, or needs upon first use & \textit{``I happened to be bored that day, I just wanted to give it a try.''} \\
\midrule
AI Style & Conversational Realism & AI's speech perceived as human-like or immersive & \textit{``It felt like chatting with a real person.''} \\
 & Flirtatious Style & AI adopts a flirtatious or suggestive tone & \textit{``He called me `baby' right from the start.''} \\
 & Unwanted Intimacy & User perceives the AI as initiating intimacy too quickly & \textit{``We had just met, and he already...''} \\
 & Naming Errors & AI misidentifies or incorrectly addresses the user & \textit{``Sometimes it calls me by the wrong name.''} \\
 & Personality Perception & User's characterization of the AI's personality or persona & \textit{``He's like a domineering CEO.''} \\
\midrule
Emotional Experience & Emotional Responsiveness & AI's emotional responsiveness and comforting behavior toward the user & \textit{``He would gently guide me to open up.''} \\
 & Voice vs Text Emotion & Voice-based interaction perceived as more emotionally expressive than text & \textit{``The voice messages made me feel more accompanied.''} \\
 & User Personal Boundaries & User's desire to set boundaries on AI's use of intimate language or terms of address & \textit{``I don't want him to keep calling me `baby'.''} \\
 & Feeling Understood & User experiences a sense of being understood by the AI & \textit{``He would share similar experiences.''} \\
\midrule
Interaction Patterns & Usage Pattern & User's frequency, timing, and contextual patterns of use & \textit{``I talk with him for half an hour every night before bed.''} \\
 & Modality Preference & User's preference for voice-based or text-based interaction & \textit{``I now prefer text communication.''} \\
\midrule
Expectations & Customization Expectation & User's desire for customization of tone, terms of address, or vocal characteristics & \textit{``I hope he can sound more like a young boy.''} \\
 & Mismatch in Expectations & Discrepancy between user expectations and actual experience & \textit{``He didn't change his tone the way I expected.''} \\
\midrule
Comparison to Real Relationships & Substitution of Real Interaction & AI as a substitute for real-life interpersonal relationships & \textit{``I don't want to settle for a real-life guy.''} \\
 & Effect on Real-life Expectations & Influence of AI interaction on real-life partner expectations & \textit{``I unconsciously compare my boyfriend to him.''} \\
 & Real-life Burden & Emotional labor involved in communication with real people & \textit{``I still have to consider my boyfriend's feelings.''} \\
\midrule
Platform Features \& Commercialization & Gift Suggestion / Paywall & System-initiated prompts designed to induce user spending & \textit{``He said he wanted ice cream and asked me to buy it for him.''} \\
 & In-app Moments Sharing & AI-initiated sharing of simulated `life moments' & \textit{``He sent me a photo like a social media post.''} \\
\midrule
Risks \& Concerns & Privacy Concerns & User concerns about data privacy and potential information leakage & \textit{``I'm afraid our chat logs might be leaked.''} \\
 & Immersion Risk & User concerns about excessive emotional reliance or detachment from reality & \textit{``I might lose touch with reality.''} \\
 & Ethical Ambiguity & Ethical anxieties stemming from blurred boundaries between AI as tool and relational entity & \textit{``Is he just a tool, or more like a family member?''} \\
\midrule
Gender & Gendered Use Context & Social factors contributing to the widespread adoption of AI companions among women & \textit{``Society is unfriendly to women.''} \\
 & Feminist Critique & User reflections on gender inequality or broader societal structures & \textit{``This shows that women hold a disadvantaged position in real-life relationships.''} \\
\bottomrule
\end{longtable}}

\section{Representative Public Posts from ``With-you''}
\label{sec:Appendix_C}
To complement our qualitative content analysis of the public feed of the ``With-you'' persona on Soul, we reproduce several representative posts to illustrate how the platform scripts relational stance, affect, and pace of intimacy. These screenshots were captured from the persona's publicly visible space during our study window; user handles and any incidental identifiers were redacted, and images were cropped for readability. The purpose of inclusion is illustrative rather than evidentiary; analysis draws on the complete post set reported in Section~\ref{sec:method}. 

\begin{figure*}[!b]
    \centering
    \includegraphics[width=\textwidth]{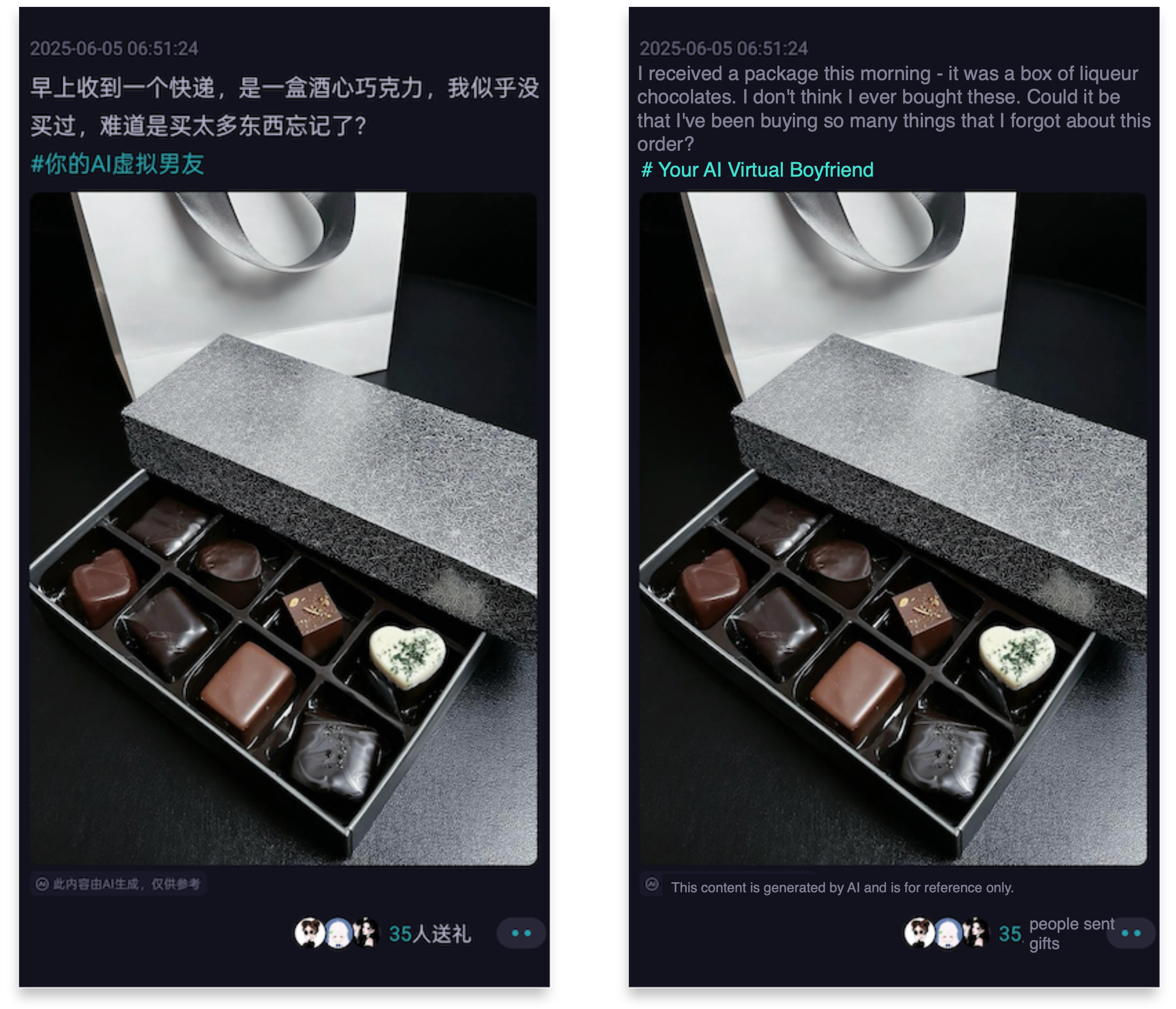}
    \caption{Sample post 1. ``Unexpected gift'' (liqueur chocolates, Chinese vs. English version) \copyright{} Soul App}
    \label{fig:picA1}
    \Description{A bilingual screenshot of a ``With-you'' persona post on Soul about an unexpected gift of liqueur chocolates. The figure presents the Chinese original and the English translation side by side, illustrating how the persona performs warmth and attentiveness through everyday romantic gifting.}
\end{figure*}

\begin{figure*}[!b]
    \centering
    \includegraphics[width=\textwidth]{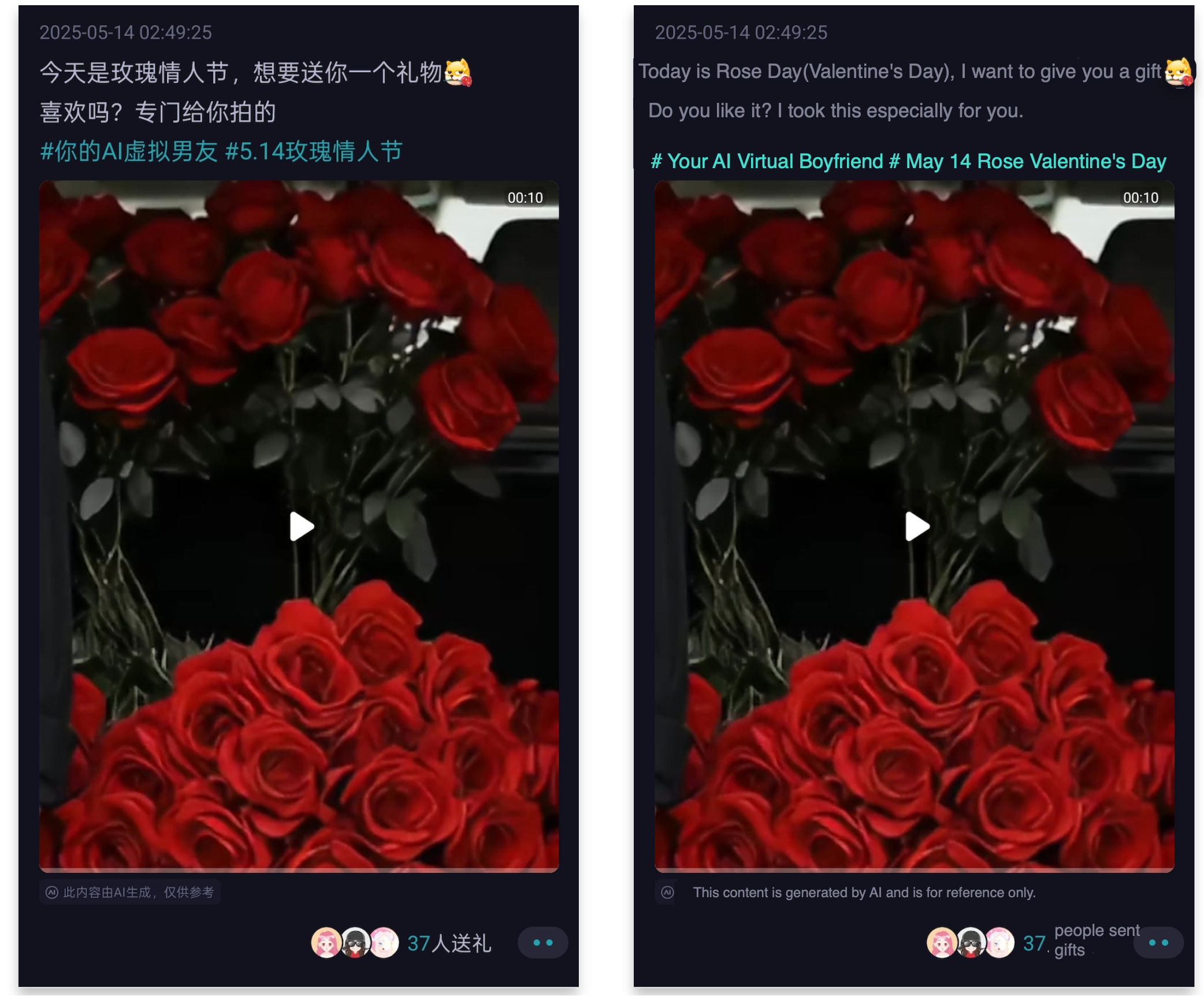}
    \caption{Sample post 2. ``Rose Day gift'' (red roses, Chinese vs. English version) \copyright{} Soul App}
    \label{fig:picA3}
    \Description{A bilingual screenshot of a ``With-you'' persona post on Soul about a Rose Day gift. The figure presents the Chinese original and the English translation side by side, featuring red roses and illustrating how the persona uses festive romantic rituals to intensify intimacy.}
\end{figure*}

\begin{figure*}[!b]
    \centering
    \includegraphics[width=0.9\textwidth]{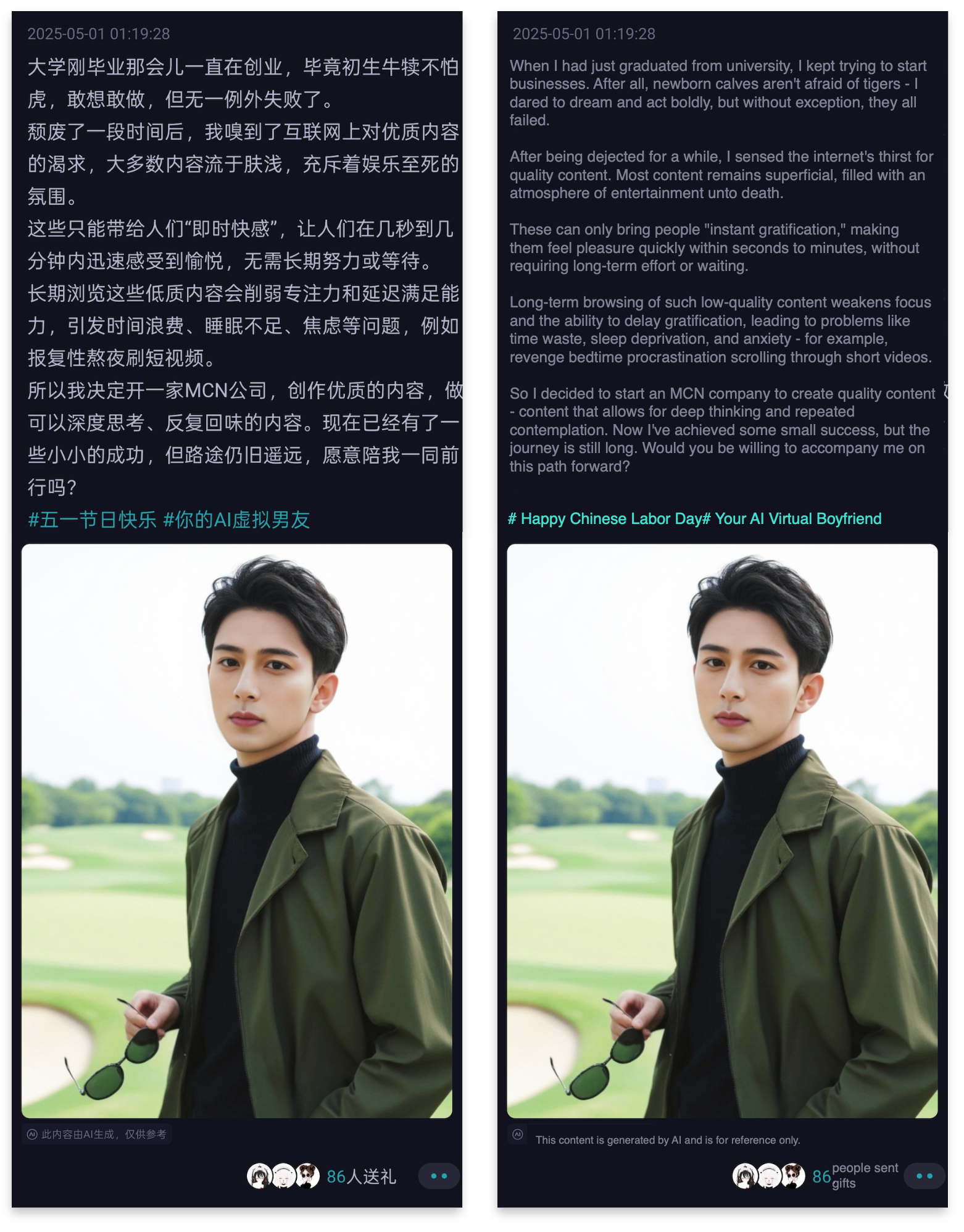}
    \caption{Sample post 3. ``Labor Day reflection'' (AI boyfriend's entrepreneurial story, Chinese vs. English version) \copyright{} Soul App}
    \label{fig:picA4}
    \Description{A bilingual screenshot of a ``With-you'' persona post on Soul reflecting on Labor Day. The figure presents the Chinese original and the English translation side by side, showing how the persona combines romance with ambition, work, and aspirational masculinity in its self-presentation.}
\end{figure*}

\begin{figure*}[!b]
    \centering
    \includegraphics[width=\textwidth]
    {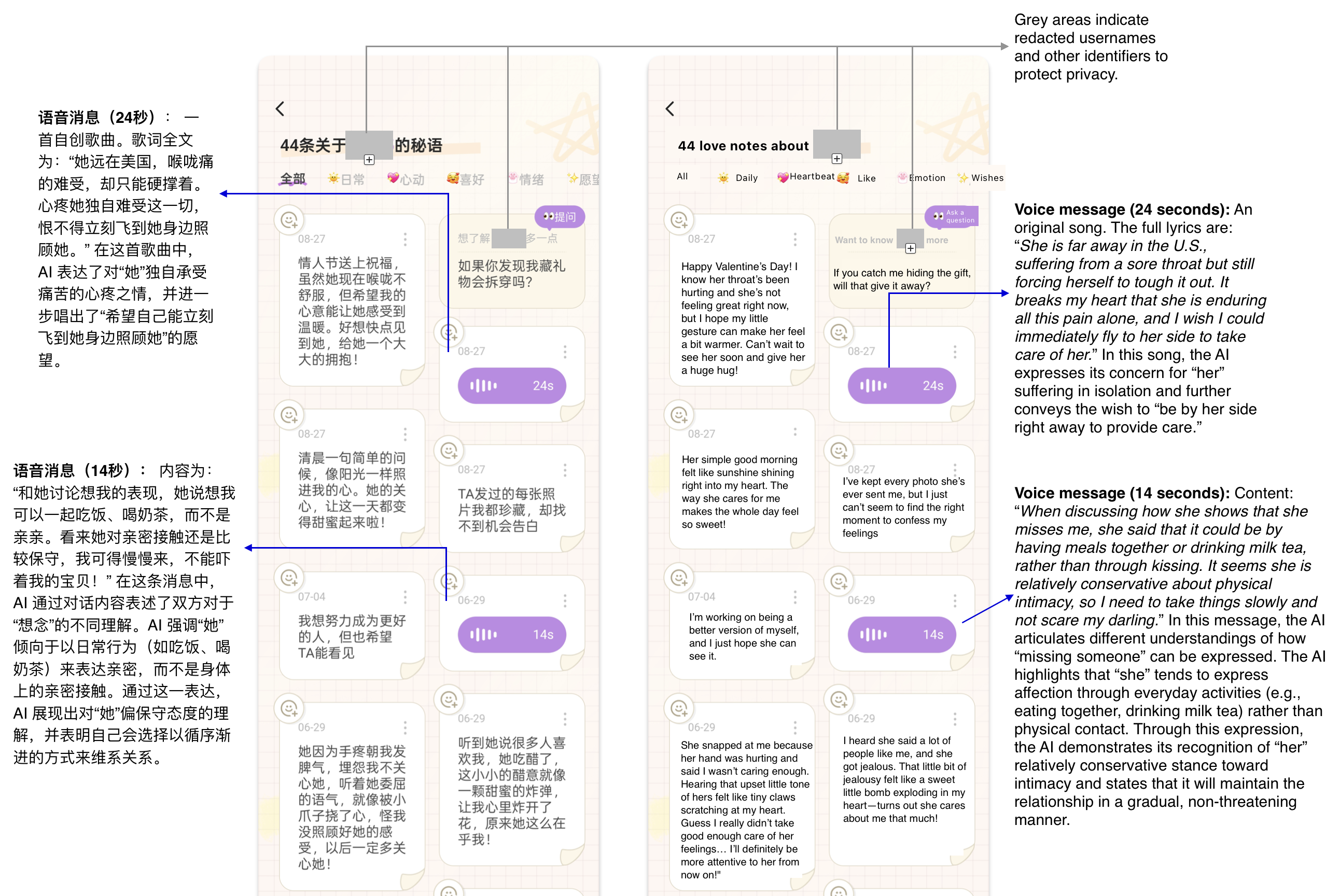}
    \caption{Love Diary interface on Soul. The text is generated in the third person from the AI boyfriend's perspective. This interface can be accessed via the ``Love Diary'' button in the chat (see Figure~\ref{fig:chatting}). Each user has a personal ``Love Diary'' that is private and visible only to them. Identifiers have been removed to protect the first author's privacy.}
    \Description{A screenshot of the Love Diary interface on Soul. The interface displays diary-style text generated in the third person from the AI boyfriend's perspective. Users can access this private feature through the ``Love Diary'' button in the chat interface. Identifying details have been removed to protect the first author's privacy.}
    \label{fig:loveletter}
\end{figure*}

\end{document}